\documentclass{aa}
\usepackage{graphicx,txfonts,natbib}
\bibpunct{(}{)}{;}{a}{}{,}

\begin{document}
\title{Corotation torques experienced by planets embedded in weakly magnetized turbulent discs}
\author{
C. Baruteau\inst{1,2}, 
S. Fromang\inst{3}, 
R. P. Nelson\inst{4},
\and 
F. Masset\inst{5,3}
}
\institute{
DAMTP, University of Cambridge, Wilberforce Road, Cambridge CB30WA, United Kingdom
\and
Department of Astronomy and Astrophysics, University of California, Santa Cruz, CA 95064, USA
 \and
 Laboratoire AIM, CEA/DSM-CNRS-Universit{\'e} Paris Diderot, IRFU/SAp, CEA/Saclay, 91191 Gif-sur-Yvette, France
 \and
 Astronomy Unit, Queen Mary, University of London, Mile End Road, London E1 4NS, United Kingdom
 \and
 Instituto de Ciencias F{\'i}sicas, Universidad Nacional Aut{\'o}noma de M{\'e}xico, Apdo. Postal 48-3, 62251-Cuernavaca, Morelos, M{\'e}xico
 }
\authorrunning{C. Baruteau et al.}
\titlerunning{Corotation torque in weakly magnetized turbulent discs}
\abstract
{The migration of low-mass planets, or type I migration, is driven by
  the differential Lindblad torque and the corotation torque in
  non-magnetic viscous models of protoplanetary discs. The corotation
  torque has recently received detailed attention, because of its
  ability to slow down, stall, or reverse type I migration. In laminar
  viscous disc models, the long-term evolution of the corotation
  torque is intimately related to viscous and thermal diffusion
  processes in the planet's horseshoe region. It is unclear how the
  corotation torque behaves in turbulent discs, and whether its
  amplitude is correctly predicted by viscous disc models.}
{This paper is aimed at examining the properties of the corotation
  torque in discs where magnetohydrodynamic (MHD) turbulence develops
  as a result of the magnetorotational instability (MRI), considering
  a weak initial toroidal magnetic field.}
{We present results of 3D MHD simulations carried out with two
  different codes. Non-ideal MHD effects and the disc's vertical
  stratification are neglected, and locally isothermal disc models are
  considered. The running time-averaged tidal torque exerted by the
  disc on a fixed planet is evaluated in three different disc models.}
{We first present simulation results with an inner disc cavity (planet
  trap). As in viscous disc models, the planet is found to experience
  a positive running time-averaged torque over several hundred orbits,
  which highlights the existence of an unsaturated corotation torque
  maintained in the long term in MHD turbulent discs. Two disc models
  with initial power-law density and temperature profiles are also
  adopted, in which the time-averaged tidal torque is found to be in
  decent agreement with its counterpart in laminar viscous disc models
  with similar viscosity alpha parameter at the planet location.
  Detailed analysis of the averaged torque density distributions
  indicates that the differential Lindblad torque takes very similar
  values in MHD turbulent and laminar viscous discs, and there exists
  an unsaturated corotation torque in MHD turbulent discs. This
  analysis also reveals the existence of an additional corotation
  torque in weakly magnetized discs.}
{Our results of 3D MHD simulations demonstrate the existence of
  horseshoe dynamics and an unsaturated corotation torque in weakly
  magnetized discs with fully developed MHD turbulence.}

\keywords{accretion discs --- magnetohydrodynamics (MHD) ---
  turbulence --- methods: numerical --- planetary systems: planet-disc
  interactions --- planetary systems: protoplanetary discs}

\maketitle

\section{Introduction}
From the early stages of their formation, planets tidally interact
with their natal protoplanetary disc. The tidal torque exerted by the
disc on a planet drives the planet's radial migration. Several
migration regimes are usually distinguished, depending on the planet's
ability to clear a gap around its orbit. This ability varies with the
planet-to-primary mass ratio, the disc aspect ratio (that is, the disc
pressure scale height-to-radius ratio), and the disc's turbulent
viscosity \citep[for more details, see][]{crida06}. The migration of
planets that do not open a gap is referred to as type I migration. It
typically applies to low-mass planets, up to a few Earth masses if the
central object has a solar mass. In massive protoplanetary discs,
intermediate-mass planets building up a partial gap (e.g., Saturn-like
planets) may undergo rapid runaway migration \citep{mp03}. More
massive planets open a clean gap and are subject to type II migration
\citep{lp86, cm07}.

The reproduction of the mass-period diagram of known exoplanets by
models of planet population synthesis is very sensitive to the
expression for the tidal torque driving type I migration
\citep[e.g.,][]{IdaLin4, IdaLin5, sli09, Mordasini09b}. The latter has
been intensively studied in two-dimensional (2D) non-magnetized,
laminar viscous disc models. In such models, the torque comprises the
differential Lindblad torque and the corotation torque. The
differential Lindblad torque corresponds to the angular momentum
carried away by the spiral density waves the planet generates in the
disc, where the flow relative to the planet becomes approximately
supersonic. In discs with decreasing temperature profiles, it is a
negative quantity \citep{tanaka2002, pbck10}, which, by itself, would
drive type I migration on timescales shorter than a few $\times 10^5$
yrs \citep[e.g.,][]{w97}. Local variations in the disc's temperature
and/or density profiles, due for example to opacity transitions
\citep{mg2004} or to dust heating \citep{hp10} may however change the
sign and magnitude of the differential Lindblad torque.

The corotation torque accounts for the exchange of angular momentum
between the planet and the coorbital flow. It has two components. One
scales with the gradient of disc vortensity (vorticity to surface
density ratio, projected along the vertical direction) across the
horseshoe region \citep{wlpi91, masset01}. The other features the
entropy gradient across the horseshoe region \citep{bm08a, pp08, mc10,
  pbck10}. The magnitude of the corotation torque depends on the
timescales for viscous and thermal diffusion inside the horseshoe
region, required to maintain the local gradients of vortensity and
entropy. In the absence of diffusion processes, both gradients would
be progressively flattened out by advection inside the horseshoe
region, and the corotation torque would ultimately vanish. This is
called saturation of the corotation torque. It may be avoided by
requiring that the viscous and thermal diffusion timescales across the
horseshoe region be shorter than the horseshoe libration timescale
\citep{pbk11,mc10}. When both diffusion timescales exceed the
horseshoe U-turn timescale (a fraction of the libration timescale),
the corotation torque is a non-linear process also known as horseshoe
drag. When they become comparable to, or shorter than the U-turn
timescale, the corotation torque decreases towards its value predicted
by linear theory \citep{pp09a}, referred to as the linear corotation
torque.

Depending on the disc gradients and diffusion processes at the planet
orbit, the amplitude of the corotation torque can be comparable to, or
larger than that of the differential Lindblad torque. The impact of
the corotation torque on type I migration can therefore be of
considerable importance. A lot of efforts have been recently put
forward to establish simple and accurate expressions for the
corotation torque as well as for the differential Lindblad torque,
which can be used in population synthesis models
\citep{mc10,pbk11}. These expressions were obtained for 2D
non-magnetic viscous disc models.

The tidal torque has also been investigated in magnetized
non-turbulent discs. The case of a 2D disc with a toroidal magnetic
field has been studied by \cite{Terquem03} through a linear
analysis. She found that the differential Lindblad torque is reduced
with respect to non-magnetized discs (waves propagate outside the
Lindblad resonances at the magneto-sonic speed rather than the sound
speed), and that there is no linear corotation torque. Instead,
angular momentum is taken away from the planet by slow MHD waves
propagating in a narrow annulus near magnetic resonances. These
results were confirmed by \cite{Fromangetal05} with non-linear 2D MHD
simulations in the regime of strong magnetic field (the plasma
$\beta$-parameter was taken equal to 2 in their study, where $\beta$
is the ratio of the thermal to magnetic pressures).  More recently, 2D
and 3D disc models with a poloidal magnetic field were investigated by
\cite{Muto08} in the shearing sheet approximation.

It is unclear how turbulence affects the tidal torque. So far, only a
couple of studies have investigated this issue. \cite{np2004}
performed 3D simulations of locally isothermal discs fully invaded by
MHD turbulence due to the MRI. They found that the running time
averaged tidal torque on a fixed protoplanet experiences rather
large-amplitude oscillations, and its final value quite substantially
differs from the torque value expected in laminar discs. Similar
results were obtained by \cite{nelson05}, who allowed the planet orbit
to evolve.  A primary reason for the observed difference between the
laminar torque and the time-averaged turbulent torque is that the 3D
MHD simulations were not converged in time. \cite{bl10} considered 2D
isothermal discs subject to stochastic forcing, using the turbulence
model originally developed by \cite{lsa04}. They showed that, when
time-averaged over a sufficient long time period, both the
differential Lindblad torque and the corotation torque behave very
similarly as in laminar viscous disc models. \cite{Uribe11} have
recently performed 3D MHD simulations of planet migration in weakly
magnetized turbulent discs, including vertical stratification and a
range of planet masses. For type I-migrating planets, they obtained a
positive running time-averaged torque in disc models where the density
near the planet's orbital radius becomes a slightly increasing
function of radius due to the disc structuring by turbulence,
suggesting the existence of an unsaturated corotation torque in such
disc models.

In this paper, we revisit the properties of the tidal torque
experienced by a fixed planet in disc models with hydromagnetic
turbulence generated by the non-linear development of the MRI.
Locally isothermal discs without vertical stratification are
considered, and non-ideal MHD effects are neglected. 3D MHD
simulations using two different codes have been carried out, using
initial values of the plasma $\beta-$parameter larger than or equal to
50. The physical model and numerical setup used in our simulations are
described in Sect.~\ref{sec:model}. In Sect.~\ref{sec:cavity} we
present results of simulations with an inner disc cavity. In viscous
disc models, an increasing density profile yields a large positive
corotation torque, which may stall type I migration. This is often
referred to as a planet trap \citep[e.g.,][]{masset06a}. Our results
of turbulent simulations show a positive time-averaged tidal torque
experienced by a planet fixed between the two edges of a cavity,
suggesting the existence of an unsaturated corotation torque in
MHD-turbulent discs. This existence is confirmed in disc models with
power-law density profiles, which are described in
Sect.~\ref{sec:power}. The time-averaged tidal torque obtained in such
discs is in decent agreement with the torque value predicted in 2D
non-magnetized viscous disc models. The discussion in
Sect.~\ref{sec:discu} gives more insight into the existence of
horseshoe dynamics in weakly magnetized turbulent discs, as well as a
detailed comparison between the torque density distributions obtained
in MHD-turbulent and laminar disc models. This comparison indicates
the existence of an additional corotation torque in discs with a weak
toroidal magnetic field, which will be investigated in a forthcoming
study. Conclusions and future directions are drawn in
Sect.~\ref{sec:conclu}.

\section{Model description}
\label{sec:model}
We explore the properties of the tidal torque between a planet and its
nascent protoplanetary disc, wherein turbulence is driven by the
non-linear development of the MRI. A question that needs to be
addressed is whether an unsaturated corotation torque exists and can
be maintained in the long term in such turbulent discs. For this
purpose, MHD simulations have been carried out using two different
codes, which are described in Sect.~\ref{sec:codes}. The physical
model and numerical setup used in our simulations are detailed in
Sect.~\ref{sec:setup}. The parameters common to all simulations
presented hereafter are listed in Table~\ref{tab:global}.
\begin{table}
  \caption{Disc, grid and planet parameters common to Sect.~\ref{sec:cavity} 
    (disc with an inner cavity) and Sect.~\ref{sec:power} (discs with a power-law density profile).}
\begin{tabular}{ll}
  Parameter & Value\\
  \hline
  \hline
  Radial resolution & 320 cells in $R \in [1;8]$\\
  Azimuthal resolution & 480 cells in $\varphi\in[0;\pi]$\\
  Vertical resolution & 40 cells in $z\in [-0.3;0.3]$\\
  Initial magnetic field & Toroidal, set up in $R \in [1.5;5]$\\
  Aspect ratio & $H/R = 0.1$ at $R=3$\\
  Planet mass & $M_{\rm p}=3\times 10^{-4} M_{\star}$\\
  Planet location & $R_{\rm p} = 3, \varphi_{\rm p} = \pi/2, z_{\rm p}=0$\\
  Planet's softening length & $\varepsilon = 0.2H(R_{\rm p})$\\
  Frame & Corotating with $R=R_{\rm p}$\\
  Radial boundary condition & Wave-killing zones in\\
  ~ & $R \in [1;1.5]$ and $R \in [7;8]$  
\end{tabular}
\label{tab:global}
\end{table}

\subsection{Codes}
\label{sec:codes}
We investigate the nature of the corotation torque, a process
occurring on the width of the planet's horseshoe region that, for
planets subject to type I migration, is smaller than the pressure
scale height \citep[e.g.,][]{mak2006}. At this scale, the turbulence
properties can potentially be influenced by details of the numerical
scheme. Thus, we use two different codes that have different
dissipation properties. This will help constrain the robustness of our
results, and their limitations.

The two codes that we use to perform the 3D MHD simulations presented
in this paper are NIRVANA \citep{nirvana1} and RAMSES
\citep{teyssier02,fromangetal06}. NIRVANA uses an algorithm very
similar to the ZEUS code to solve the equations of ideal MHD
\citep{zeus}. RAMSES is a finite-volume code that uses the
MUSCL-Hancok Godunov scheme. For the purpose of this project, a
uniform grid version of the code (i.e., without any Adaptive Mesh
Refinement capabilities) was extended to allow the use of cylindrical
coordinates.  For comparison, results of 2D hydrodynamical simulations
are also presented, performed with the codes RAMSES and FARGO
\citep{fargo1, fargo2}. (None of these 2D simulations includes a
magnetic field). FARGO is very similar to the 2D hydrodynamical
version of NIRVANA. Its specificity is to use a change of rotating
frame on each ring of the grid, which increases the timestep
significantly.

Results of simulations are expressed in the following units: the mass
unit is the mass of the central star ($M_{\star}$), the length unit is
the planet's (fixed) orbital separation ($R_{\rm p}$), and the time
unit is the planet's orbital period ($T_{\rm orb}$) divided by
$2\pi$. Whenever time is expressed in orbits, it refers to the orbital
period at the planet location.

\subsection{Physical model and numerical setup}
\label{sec:setup}

\subsubsection{Gas disc model}
\label{sec:physmodel}
We adopt a simple 3D magnetized disc model in which non-ideal MHD
effects, self-gravity and vertical stratification are neglected. No
explicit kinematic viscosity is included.  The ideal MHD equations are
solved in a cylindrical coordinate system $\{R, \varphi, z\}$ centred
on the central star, with $R\in [1;8]$, $\varphi\in[0;\pi]$ and $z\in
[-0.3;0.3]$. The frame rotates with angular frequency equal to the
Keplerian frequency at $R=3$ (the fixed location of the planet, when
included), and the indirect term that accounts for the acceleration of
the central star by the planet is included in the equations of motion
for those simulations that include a planet.  For simplicity, a
locally isothermal equation of state is used, with the gas pressure
$P_{\rm gas}$ and mass volume density $\rho$ satisfying $P_{\rm gas} =
\rho c^2_{\rm s}$, where the gas sound speed $c_{\rm s}$ is specified
as a fixed function of radius. It is related to the disc's pressure
scale height $H$ through $H = c_{\rm s} / \Omega_{\rm K}$, with
$\Omega_{\rm K}$ the local Keplerian angular velocity. The disc aspect
ratio, $h=H/R$, is defined to be 0.1 at $R=3$. The vertical extent of
the disc model is thus equal to two pressure scale heights at the
planet's orbital radius. Ideally we would prefer to choose a smaller
disc thickness, more typical of protoplanetary disc models.  But the
resolution requirement of maintaining a turbulent flow driven by the
MRI, combined with the need to run simulations for many hundreds of
planet orbits, means that we are forced to adopt a relatively thick
disc model.

The disc is set up in radial equilibrium, with the centrifugal
acceleration and the radial acceleration related to the pressure
gradient balancing the gravitational acceleration due to the central
star (function of $R$ only). A small level of random noise ($5\%$ of
the local sound speed) is added to each component of the gas
velocity. The initial radial profiles of the disc's density and
temperature will be specified in Sects.~\ref{sec:cavity}
and~\ref{sec:power}. The initial magnetic field is purely toroidal
with a net azimuthal flux. Its radial profile is set by the choice for
the plasma $\beta$-parameter, which is taken to be constant throughout
the disc region in which the initial field is introduced.  Several
values of $\beta$ in the range $[50-400]$ are considered throughout
this study, depending on the model. Their value will be specified
below.  The initial magnetic field is introduced everywhere except
near the radial boundaries, where the field is set to zero in $R \in
[1;1.5]$ and $R \in [5;8]$.

The long term action of the turbulent stresses causes the disc density
profiles to evolve significantly, leading eventually to most of the
mass being relocated to the close vicinity of the inner and outer
radial boundaries. To avoid this and obtain a steady-state density
profile, we adopt an approach similar to that of \cite{ng10}, and
reinforce the original density profile on some specified time
scale. This procedure, which we call the mass-adding procedure for
future reference, is achieved by solving
\begin{equation}
  \frac{\partial\rho}{\partial t} = -\frac{\langle\rho\rangle_{\varphi} - \rho_0}{\tau}
\label{eq:addmass}
\end{equation}
alongside the ideal MHD equations. In Eq.~(\ref{eq:addmass}),
$\langle\rho\rangle_{\varphi}$ denotes the azimuthally-averaged
density (function of $R$ and $z$), $ \rho_0$ is the initial density
(function of $R$ only), and $\tau$ is set to 20 local orbital periods.
We checked by means of 2D hydrodynamical viscous simulations with a
kinematic viscosity corresponding to a similar amount of turbulence as
in our MHD simulations that the mass-adding procedure has a very small
impact on the time evolution of the tidal torque, since the planet we
consider (see Sect.~\ref{sec:planetparam}) does not open a gap over
the duration of our simulations. Some additional remarks on the use of
the mass-adding procedure will be made in Sect.~\ref{sec:addmass}.

\subsubsection{Planet parameters}
\label{sec:planetparam}
We adopt a planet of mass $M_{\rm p} = 3 \times 10^{-4} M_{\star}$,
which is held on a fixed circular orbit at $R_{\rm p}=3$, $z_{\rm
  p}=0$ (here and in the following, all quantities with the subscript
p are meant to be evaluated at the planet's orbital radius). This
corresponds to a Saturn-mass planet if the central object is a
Sun-like star. Given the large disc aspect ratio in our model ($h_{\rm
  p}=0.1$), the dimensionless parameter determining the flow
non-linearity near the planet $M_{\rm p} / (h_{\rm p}^3 M_{\star}) =
0.3 \lesssim 1$. In addition, the large turbulent stress generated by
MHD turbulence (the effective viscosity parameter $\alpha$ is a few
percent, as we shall see in the next sections) prevents the planet
from opening a gap over the duration of our calculations (several
hundred orbits). The planet mass we consider is therefore relevant to
the type I migration regime, although it is somewhat larger than would
normally be adopted in studies of the migration of low-mass
planets. The need to obtain converged estimates of the torque
experienced by the planet over runs times of several hundred planet
orbits, however, has led us to adopt this larger than preferred value.
The gravitational potential of the planet is independent of $z$, this
being consistent with our cylindrical disc setup. It is smoothed over
a softening length, $\varepsilon$, equal to $0.2 H(R_{\rm p})$. The
calculation of the force exerted by the disc on the planet excludes
gas located within the planet's Hill radius ($R_{\rm H}$, the excluded
volume is a cylinder of radius $R_{\rm H}$ centred around the planet's
position).
  
\subsubsection{Grid resolution and boundary conditions}
\label{sec:resol}
\par\noindent\emph{Resolution---}
The grid resolution in all the simulations presented in
Sects.~\ref{sec:cavity} and~\ref{sec:power} is $(N_R, N_{\varphi},
N_z) = (320, 480, 40)$.  At the planet location, the pressure scale
height is thus resolved by about 15 cells along the radial and
azimuthal directions, and by 20 cells along the vertical
direction. Since we aim to investigate the properties of the
corotation torque, the radial extent of the planet's horseshoe region
needs to be sufficiently resolved. For type I migrating objects, its
half-width, $x_{\rm s}$, is $\simeq 1.2 R_{\rm p} \sqrt{M_{\rm p}/
  (h_{\rm p}\,M_{\star})}$ \citep[e.g.,][]{mak2006}. Here, it is
resolved by about 9 cells along the radial direction, which is similar
to the resolution used in recent studies of the corotation torque
\citep[e.g.,][]{bl10}.
\\
\par\noindent\emph{Azimuthal domain---}
\begin{figure}
  \includegraphics[width=\hsize]{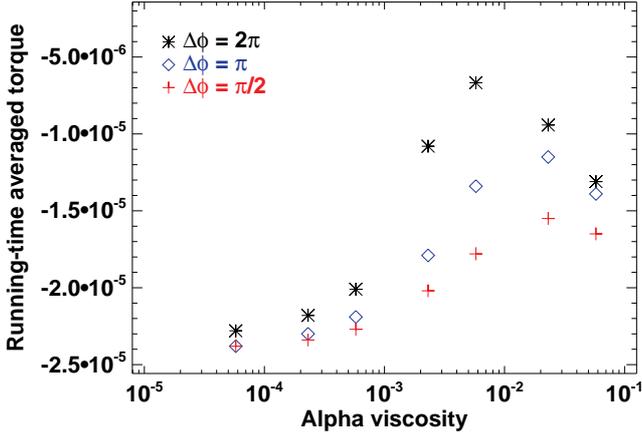}
  \caption{\label{fig:extent}Results of 2D viscous disc models with
    FARGO, using a constant kinematic viscosity. The steady running
    time-averaged specific torque on the planet (code units) is
    plotted against the alpha viscosity at the planet
    location. Results are obtained for several values of the grid's
    azimuthal extent $\Delta\varphi$ but with the same resolution.}
\end{figure}
As already mentioned in Sect.~\ref{sec:physmodel}, our 3D MHD
simulations cover an azimuthal extent $\Delta\varphi = \pi$. This
choice is the best compromise between several limitations. Having
$\Delta\varphi = 2\pi$ is a natural choice for studying the corotation
torque, and to our knowledge all works on the corotation torque have
used this azimuthal extent. But, at fixed resolution, this choice
requires a maximally large $N_{\varphi}$, the number of grid cells
along the azimuthal direction. At the same resolution, decreasing
$\Delta\varphi$ reduces $N_{\varphi}$, but at the cost of decreasing
the corotation torque.  This point is illustrated in
Fig.~\ref{fig:extent} for 2D hydrodynamical simulations with
FARGO. These simulations use the same setup as described above for the
3D MHD simulations, except that a constant kinematic viscosity $\nu$
is included.  The initial surface density and temperature profiles
decrease as $R^{-1/2}$ and $R^{-1}$, respectively.
Fig.~\ref{fig:extent} displays the (stationary) running time-average
of the specific torque\footnote{As is conventional, we refer to torque
  as the torque exerted by the disc on the planet. If $\Gamma$ denotes
  the torque, the running time-averaged torque, $\overline{\Gamma}$,
  is calculated as $\overline{\Gamma}(t) = t^{-1} \times \int_0^t
  \Gamma(u) du$.}  obtained for three series of runs with
$\Delta\varphi = \pi/2$, $\pi$ and $2\pi$, for which $N_{\varphi} =
240, 480$ and 960, respectively.  For each series, the alpha viscosity
parameter at the planet location, $\alpha_{\rm p} \equiv \nu /
(H^2_{\rm p}\,\Omega_{\rm K, p})$, ranges from $6\times 10^{-5}$ to
$6\times 10^{-2}$.  At a given $\alpha_{\rm p}$, the corotation torque
can be estimated as the difference between the total torque, and the
total torque at smallest $\alpha_{\rm p}$ (which is a good
approximation for the differential Lindblad torque; note that its
value remains almost unchanged by reducing $\Delta\varphi$ from $2\pi$
to $\pi/2$).  The corotation torque is approximately maximum when the
viscous diffusion timescale across the horseshoe region, $\tau_{\rm
  visc} \sim x_{\rm s}^2 / \nu$, is shorter than the libration
timescale,
\begin{equation}
  \tau_{\rm lib}(\Delta\varphi) = \frac{8\pi R_{\rm p}}{3\Omega_{\rm p}x_{\rm s}} \times \frac{\Delta\varphi}{2\pi},
\end{equation}
but exceeds the U-turn timescale \citep{bm08a},
\begin{equation}
\tau_{\rm U-turn} \approx
h_{\rm p}\times \tau_{\rm lib}(\Delta\varphi = 2\pi).
\label{eq:tauUturn}
\end{equation}
We note that, for $\Delta\varphi = 2\pi$, the value of $\alpha_{\rm
  p}$ at which the corotation torque is maximum is in good agreement
with the expression $0.16\,(M_{\rm p}/M_{\star})^{3/2}\,h_{\rm
  p}^{-4}$ of \cite{bl10}, which equals $\sim 8\times 10^{-3}$ for our
setup. Decreasing $\Delta\varphi$ reduces $\tau_{\rm lib}$ in
comparison to $\tau_{\rm U-turn}$, which implies that
\begin{enumerate}
\item[-] a proportionally larger viscosity is required to maintain the
  same saturation level of the corotation torque (that is, the same
  ratio $\tau_{\rm visc} / \tau_{\rm lib}$),
\item[-] the maximum amplitude of the corotation torque decreases as
  the ratio $\tau_{\rm lib} / \tau_{\rm U-turn}$ approaches unity. The
  linear corotation torque (that is, the corotation torque in the
  limit of very high viscosities) is not significantly modified by
  reducing $\Delta\varphi$.
\end{enumerate}
Taking $\Delta\varphi = \pi$ is found to be a good compromise between
a large enough corotation torque, easily measurable in turbulent
simulations, and a tractable numerical resolution. For comparison
purposes, the 2D hydrodynamical simulations presented in
Sects.~\ref{sec:cavity} and~\ref{sec:power} also have $\Delta\varphi =
\pi$. In the following, time is expressed in planet orbital periods.
Despite the choice for $\Delta\varphi = \pi$, the planet's orbital
period is implicitly calculated as $2\pi / \Omega_{\rm p}$.
\\
\par\noindent\emph{Boundary conditions---}
Periodic conditions are adopted at the azimuthal and vertical
boundaries. The treatment of the radial boundary condition differs in
the two codes. Hydrodynamic variable fluctuations are damped
identically, using the scheme described in \cite{valborro06}. This
scheme, which relaxes the density and velocity components towards
their initial value, is applied over $R\in [1;1.5]$ and $R\in
[7;8]$. In order to avoid the magnetic field reaching large values
near the boundaries, different strategies are used in both codes. In
NIRVANA, a reflecting boundary condition is applied to the gas radial
velocity that prevents the magnetic field from penetrating into the
ghost zones. In RAMSES, a magnetic resistivity ($\eta$) is used in the
radial buffer zones ($\eta_{\rm in}=5 \times 10^{-4}$ and $\eta_{\rm
  out}=10^{-3}$ at the inner and outer boundaries,
respectively). However, such an efficient field diffusion
simultaneously causes the toroidal magnetic flux to escape from the
grid, ultimately resulting in the volume-averaged $\alpha$ value
decreasing with time. To avoid this effect, at each timestep, an
additional axisymmetric toroidal field is added in each grid cell to
guarantee that the total initial toroidal flux is conserved. In doing
so, the additional toroidal field is radially distributed such that
the ratio of its associated magnetic pressure and of the initial
thermal pressure is uniform over the computational domain.

\section{Disc model with an inner cavity}
\label{sec:cavity}
\begin{figure}
  \resizebox{\hsize}{!}{\includegraphics{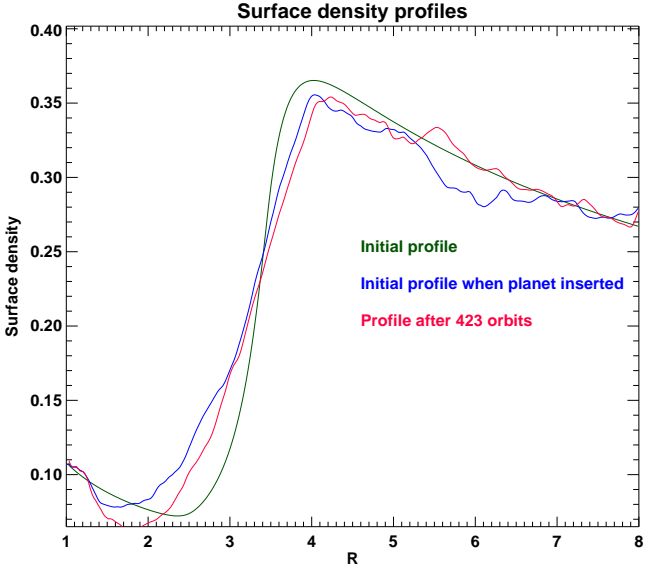}}
  \caption{\label{fig:rpn_sigmatrap} Surface density as a function of
    radius shown at different times during the simulation with an
    inner disc cavity.}
\end{figure}
In locally isothermal discs, the corotation torque is expected to be
particularly strong in regions where the density gradient is large,
and for this reason we have performed a sequence of simulations with
the planet located in a disc with an inner cavity. The planet itself
orbits in the transition region between the outer high-density disc
and the inner low-density cavity -- a region often referred to as a
planet trap \citep{masset06a, morby08}.

The initial disc model in this case was set up by first running a
laminar disc model with initial density and temperature profiles
scaling as $R^{-1/2}$ and $R^{-1}$, respectively. The functional form
for the kinematic viscosity, $\nu$, used to establish the cavity is
\begin{eqnarray}
  \nu & = & \nu_{\rm in}  \;\;\;\;\;\;\;\; \textrm{if \hspace{1mm} $R < R_{\rm in}$ } \nonumber \\
  \nu & = & \nu_{\rm in } + \frac{(\nu_{\rm out} - \nu_{\rm in})(R - R_{\rm in})}{(R_{\rm out} - R_{\rm in})} 
  \;\;\;\;\;\;\;\; \textrm{if \hspace{1mm} $R_{\rm in} \le R \le R_{\rm out}$}\\
  \nu & = & \nu_{\rm out} \;\;\;\;\;\; \textrm{if \hspace{1mm} $R > r_{\rm out}$}  \nonumber, 
\label{eqn:nu-trap}
\end{eqnarray}
with $\nu_{\rm in} = 2 \times 10^{-3}$, $\nu_{\rm out}=\nu_{\rm
  in}/7$, $R_{\rm in}=2.5$ and $R_{\rm out} = 3.5$. All other
parameters (e.g., grid resolution, boundary conditions) are as
described in Sect.~\ref{sec:model}. Note that the grid's azimuthal
extent equals $\pi$ in all the simulations presented from this section
onwards. The numerical simulations presented in the present section
were carried out with NIRVANA.

\begin{figure*}
  \resizebox{\hsize}{!}
  {
    \includegraphics{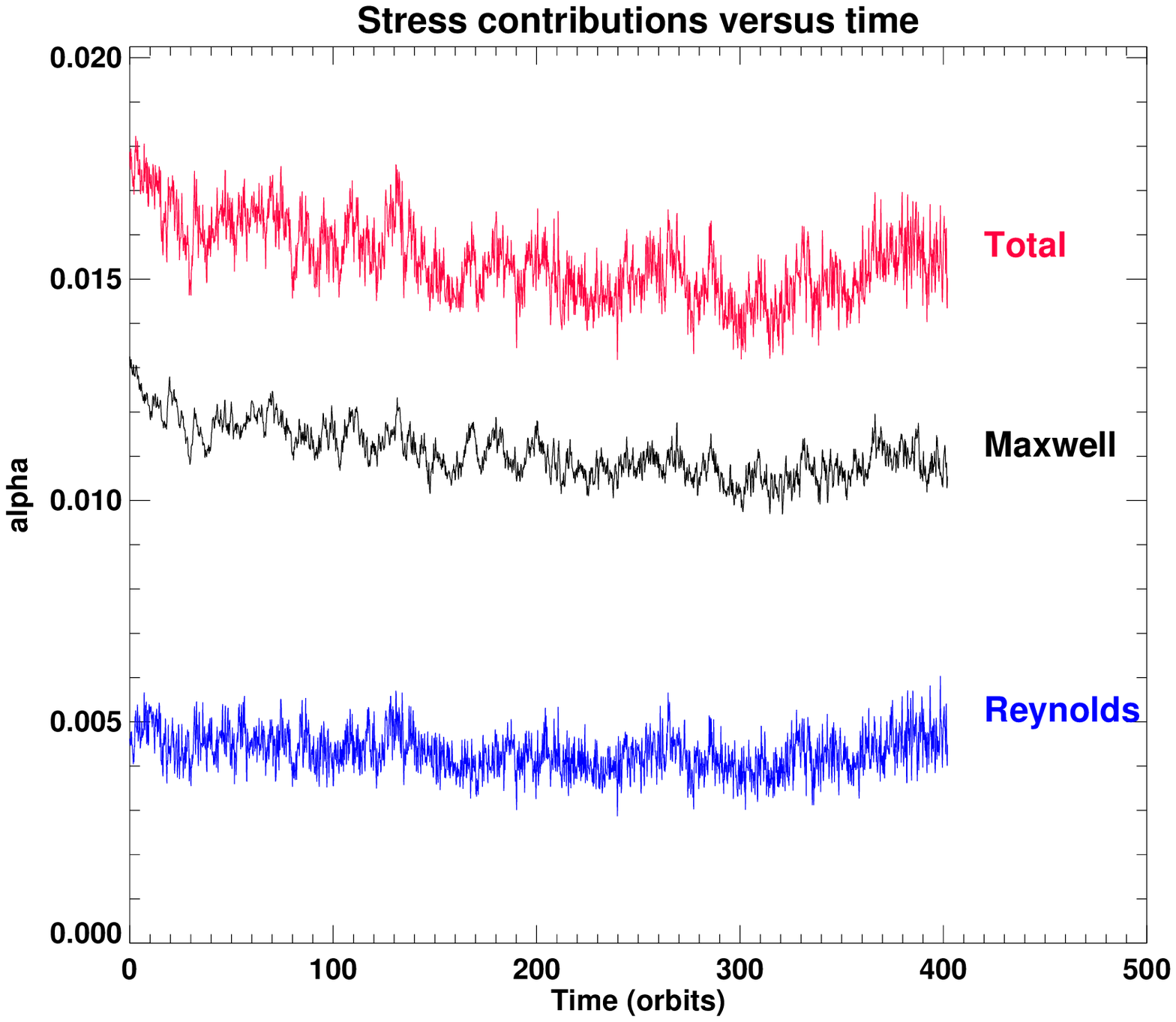}
    \includegraphics{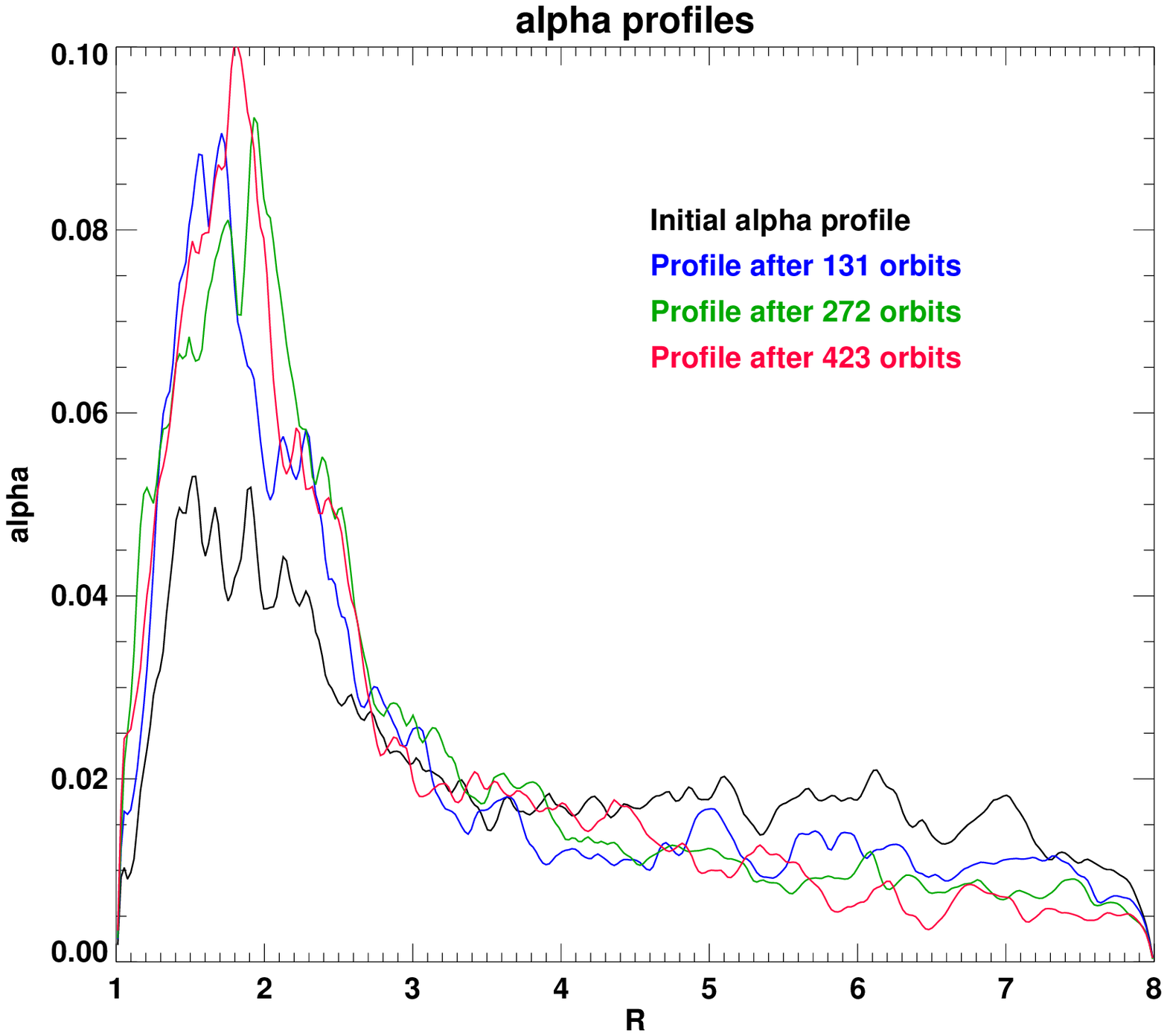}
   } 
   \caption{\label{fig:rpn_alpha}Left: volume-averaged alpha values
     for the disc model with an inner cavity, shown from the point in
     time when the planet is inserted in the disc.  As in all other
     panels, time is expressed in orbital periods at $R=3$ (the
     planet's fixed location). Right: azimuthally- and
     vertically-averaged profiles of alpha for the same disc model,
     shown at different times after the planet inclusion.}
\end{figure*}
The laminar simulation was run for $\sim 10^4$ orbits, after which
time the surface density profile had achieved the near steady-state
shown by the smooth (green) curve in Fig.~\ref{fig:rpn_sigmatrap}
(labelled as ``Initial profile"). This profile, and the associated
azimuthal velocity profile, were used to construct a 3D cylindrical
disc model into which a purely toroidal magnetic field with
$\beta=100$ was introduced throughout the disc. This magnetized disc
model was evolved for 76 orbits, until turbulence was fully developed
throughout. The mass-adding procedure described in
Sect.~\ref{sec:physmodel} was used to relax the density profile
towards the smooth initial density profile throughout the turbulent
run.

\begin{figure*}
  \resizebox{\hsize}{!}{\includegraphics{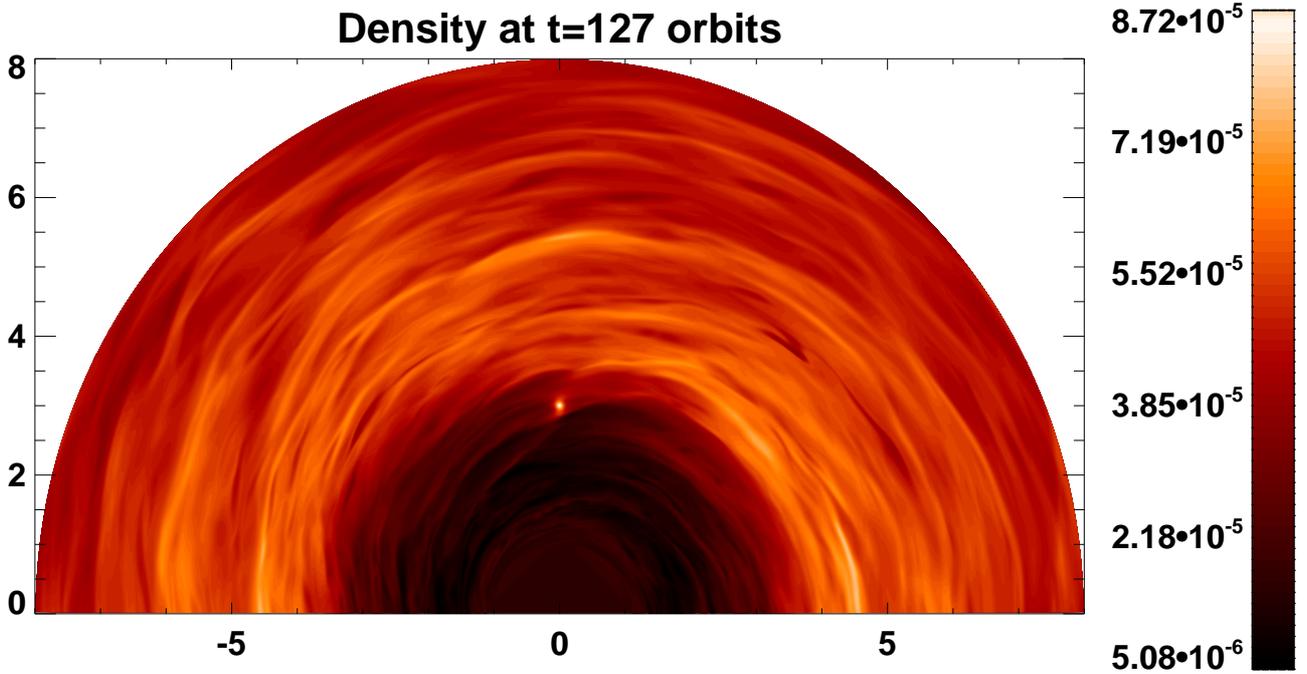}}
  \caption{\label{fig:rpn_niceplot}Contours of the gas midplane
    density in the disc model with an inner cavity, shown 127 orbits
    after the planet was inserted.}
\end{figure*}

The turbulent simulation was restarted by including a planet of mass
$M_{\rm p}=3 \times 10^{-4}M_{\star}$ at $R=3$. The disc-planet system
was evolved for $\sim 420$ orbits. The evolution of the surface
density profile (vertically-integrated density) during this time is
shown in Fig.~\ref{fig:rpn_sigmatrap}. Although there is a modest
change in the surface density gradient at the cavity edge, due to the
MHD turbulence, we see that a steep gradient is maintained. The time
evolution of the volume-averaged contributions to the effective
turbulent alpha viscosity (stress to thermal pressure ratios) is shown
in the left panel of Fig.~\ref{fig:rpn_alpha}, and it is clear that
these globally averaged stresses maintain a near
steady-state. Snapshots of the variation of the total stress as a
function of radius are shown in the right panel of
Fig.~\ref{fig:rpn_alpha}, and again a near steady-state is
observed. The effective $\alpha$ profile shows a maximum within the
cavity, due to the low density and pressure there. A contour plot
representing the disc midplane density after 127 planet orbits is
displayed in Fig.~\ref{fig:rpn_niceplot}, showing the position of the
planet sitting at the edge of the cavity.

\begin{figure}
  \resizebox{\hsize}{!}{\includegraphics{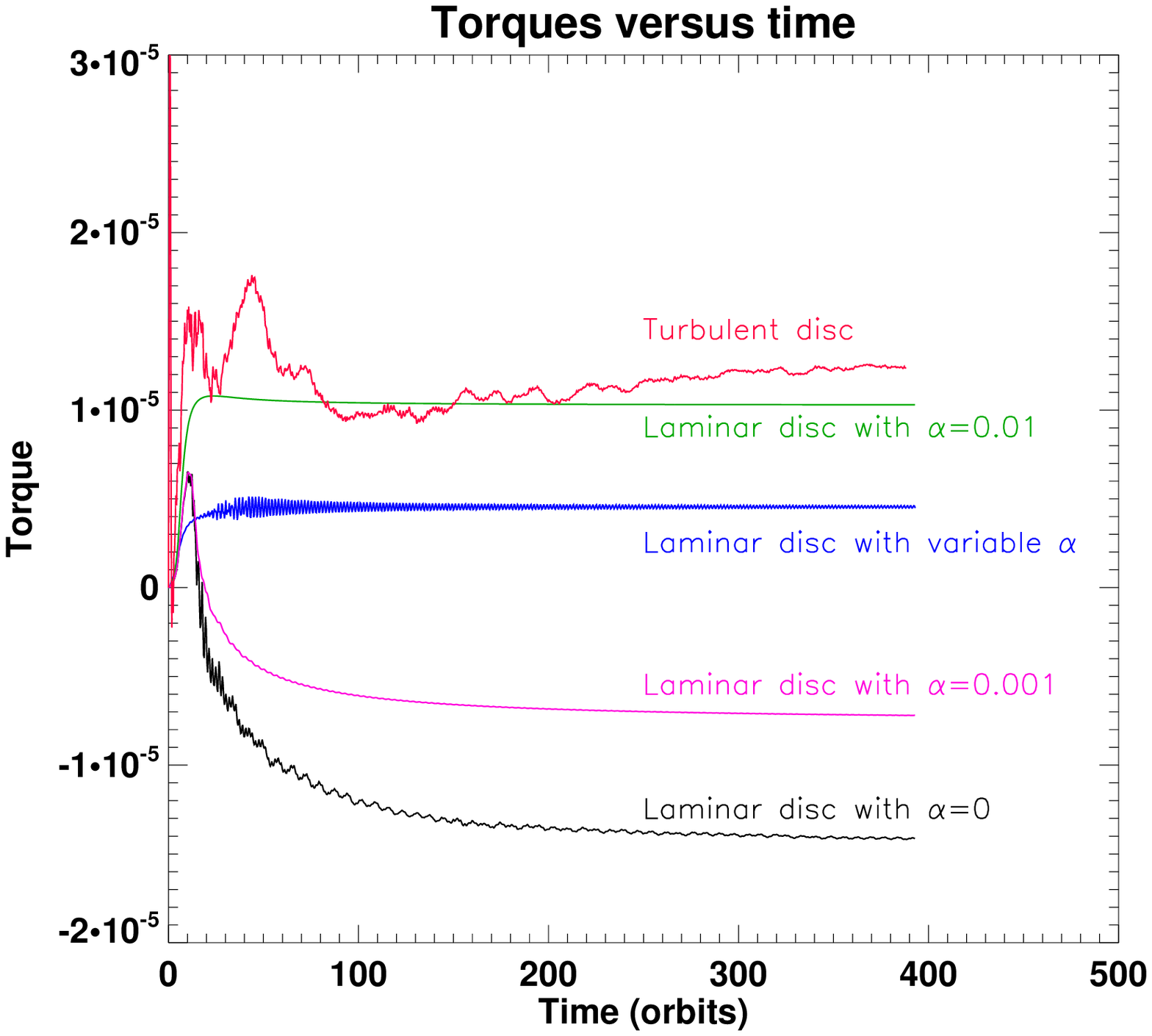}}
  \caption{\label{fig:rpn_tqtime}Running time-averaged torque obtained
    with an inner disc cavity: comparison of laminar and turbulent
    disc models.}
\end{figure}

The running time-average of the torque for the turbulent disc run is
shown by the red curve in Fig.~\ref{fig:rpn_tqtime}. It remains
positive, clearly indicating that the corotation torque is strong and
maintained in an unsaturated state throughout the simulation. For
comparison, we ran four additional 2D simulations using laminar disc
models with different values of the kinematic viscosity. The initial
surface density profile is the same as the initial profile shown by
the green curve in Fig.~\ref{fig:rpn_sigmatrap}. For comparison, the
torques obtained in the 2D simulations have been multiplied by the
scaling factor $f=0.6$ to account for the vertical extent of the 3D
simulations, where $z$ extends from -0.3 to 0.3. We see that the
laminar disc model with $\alpha_{\rm p}=0.01$ (shown by the green
curve in Fig.~\ref{fig:rpn_tqtime}) maintains a total torque whose
cumulative time-average is similar to that of the turbulent run, but
note that the surface density profiles in both runs differ about the
planet location, as can be seen in Fig.~\ref{fig:rpn_sigmatrap}.

The laminar run represented by the blue line in
Fig.~\ref{fig:rpn_tqtime} adopted the kinematic viscosity profile
given by Eq.~(\ref{eqn:nu-trap}), such that $\nu=1.14 \times 10^{-3}$
at the planet location, corresponding to the rather large value of
$\alpha_{\rm p} \sim 0.6$. The maximum unsaturated corotation torque
is expected to occur when $\alpha_{\rm p} \sim 2 \times 10^{-2}$ (see
Fig.~\ref{fig:extent}), and for values larger than this, the
corotation torque approaches a smaller value predicted by linear
theory. Our results are consistent with this. The magenta line in
Fig.~\ref{fig:rpn_tqtime} shows the result from a laminar simulation
with $\alpha_{\rm p}=10^{-3}$, a value for which we expect the
corotation torque to be partially saturated. This expectation in
confirmed by the simulations, and we see that the torque obtained with
$\alpha_{\rm p}=10^{-3}$ lies between those for $\alpha_{\rm p}=0$ and
$\alpha_{\rm p}=10^{-2}$. The black curve plotted in
Fig.~\ref{fig:rpn_tqtime} is the torque obtained for an inviscid
laminar disc. Clearly the net negative torque demonstrates that the
corotation torque in this case becomes fully saturated, leaving only
the differential Lindblad torque.

These cavity runs show that MHD turbulence with $\alpha_{\rm p} \simeq
0.02$ allows the corotation torque to be sustained at a value close to
its maximum, unsaturated value for the disc and planet parameters that
we have considered.  Significant changes in the effective $\alpha$,
due to the turbulence being more or less vigorous, however, will cause
the corotation torque to weaken. Our runs demonstrate that a planet
trap maintained in a turbulent protoplanetary disc can be effective in
preventing the large scale migration of embedded protoplanets.

As a conclusion to this section, our results strongly suggest the
existence of an unsaturated corotation torque in turbulent
protoplanetary discs with weak magnetic field (here $\beta = 100$).
Although the agreement with 2D viscous laminar simulations is quite
good, the complexity of the disc structure resulting from the presence
and time-variation of the cavity makes it difficult to perform a
quantitative analysis. The strongly increasing density profile about
the planet location, the precise slope of which significantly impacts
the corotation torque, and the fact the density profile can be
approximated as a power-law function of $R$ only over a limited region
($\lesssim \pm 3H$) about the planet location hinder a quantitative
comparison between the torques displayed in Fig.~\ref{fig:rpn_tqtime}
and analytic torque formulae. In Sect.~\ref{sec:power}, a suite of
idealized simulations using power-law density profiles are constructed
that will help study quantitatively the corotation torque properties.

\section{Disc models with a power-law density profile}
\label{sec:power}
The results of the turbulent disc model with an inner cavity presented
in Sect.~\ref{sec:cavity} strongly suggest the existence of an
unsaturated corotation torque in discs with hydromagnetic turbulence
generated by the MRI. In the present section, we investigate the
corotation torque properties in turbulent discs with initial density
profiles that are power-law functions of radius. In contrast to
Sect.~\ref{sec:cavity}, the simulations in this section have been
carried out with both NIRVANA and RAMSES. A constant initial plasma
$\beta-$parameter is adopted throughout the disc, which is equal to 50
in the runs performed with NIRVANA, and to 400 with RAMSES. As will be
shown below, these different values are found to give similar
volume-averaged alpha values in both codes.

Two disc models are considered in this section, whose parameters are
summarized in Table~\ref{tab:power}. In Model 1, the initial density
and temperature profiles, denoted by $\rho_0$ and $T_0$, decrease as
$R^{-1/2}$ and $R^{-1}$, respectively. In Model 2, $\rho_0 \propto
R^{-3/2}$ and $T_0$ is uniform. In both models, we use $\rho_0(R=1) =
10^{-4}$ in code units.  Assuming a Sun-like star, and that the
star--planet separation is 5 AU, this corresponds to an initial
surface density at the planet location $\simeq 180$ g cm$^{-2}$ in
Model 1, and $\simeq 60$ g cm$^{-2}$ in Model 2. All other parameters
are those described in Sect.~\ref{sec:model}. Recall in particular
that $T_0(R=R_{\rm p})$ is chosen such that $h_{\rm p} = 0.1$. The
results of both models are described in Sects.~\ref{sec:model1}
and~\ref{sec:model2}, respectively. In the absence of magnetic field
and turbulence, the corotation torque would vanish in Model 2, but not
in Model 1. However, we do not know whether and how magnetic effects
modify the dependence of the corotation torque with density and
temperature gradients. The tidal torque values obtained in both models
can at best hint whether MHD turbulence impacts the differential
Lindblad torque and the corotation torque, but they cannot allow us to
work out quantitatively how each torque is altered by MHD turbulence.
\begin{table}
  \caption{Disc parameters of the turbulent runs in Sect.~\ref{sec:power}, 
    where initial power-law density profiles are adopted. Results obtained   
    with Model 1 and Model 2 are presented in Sect.~\ref{sec:model1} 
    and~\ref{sec:model2}, respectively.}
\begin{tabular}{ll}
  Parameter & Value\\
  \hline
  \hline
  Initial gas density & $\rho_0 \propto R^{-1/2}$ (Model 1)\\
  ~ & $\rho_0 \propto R^{-3/2}$ (Model 2)\\
  Initial gas temperature & $T_0 \propto R^{-1}$ (Model 1)\\
  ~ & $T_0$ uniform (Model 2)\\
  Plasma $\beta-$parameter & $\beta = 50$ (NIRVANA), $\beta = 400$ (RAMSES)
\end{tabular}
\label{tab:power}
\end{table}

\subsection{Model 1: initial density $\propto R^{-1/2}$ and
  temperature $\propto R^{-1}$}
\label{sec:model1}
\begin{figure*}
  \resizebox{\hsize}{!}
  {
    \includegraphics{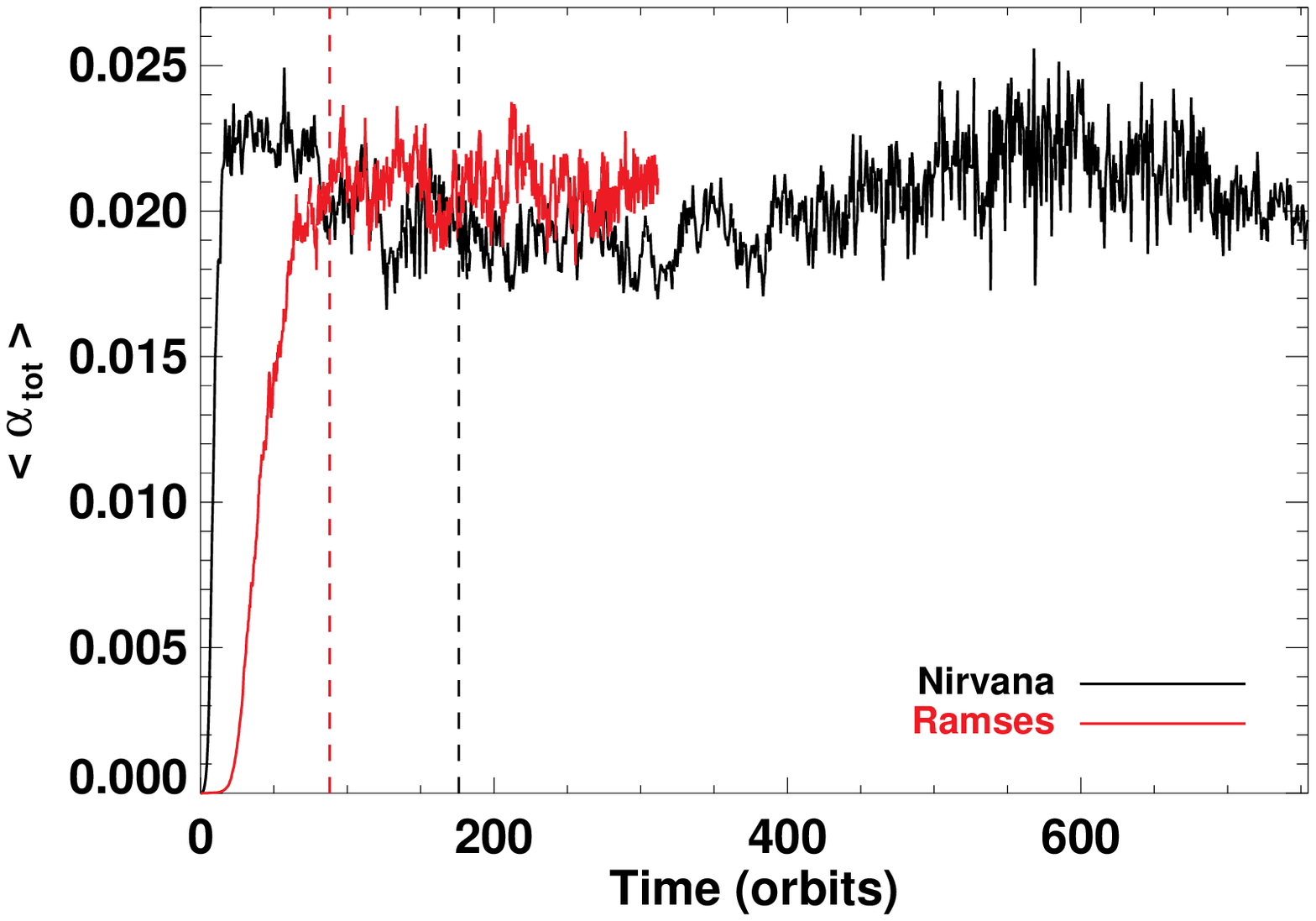}
    \includegraphics{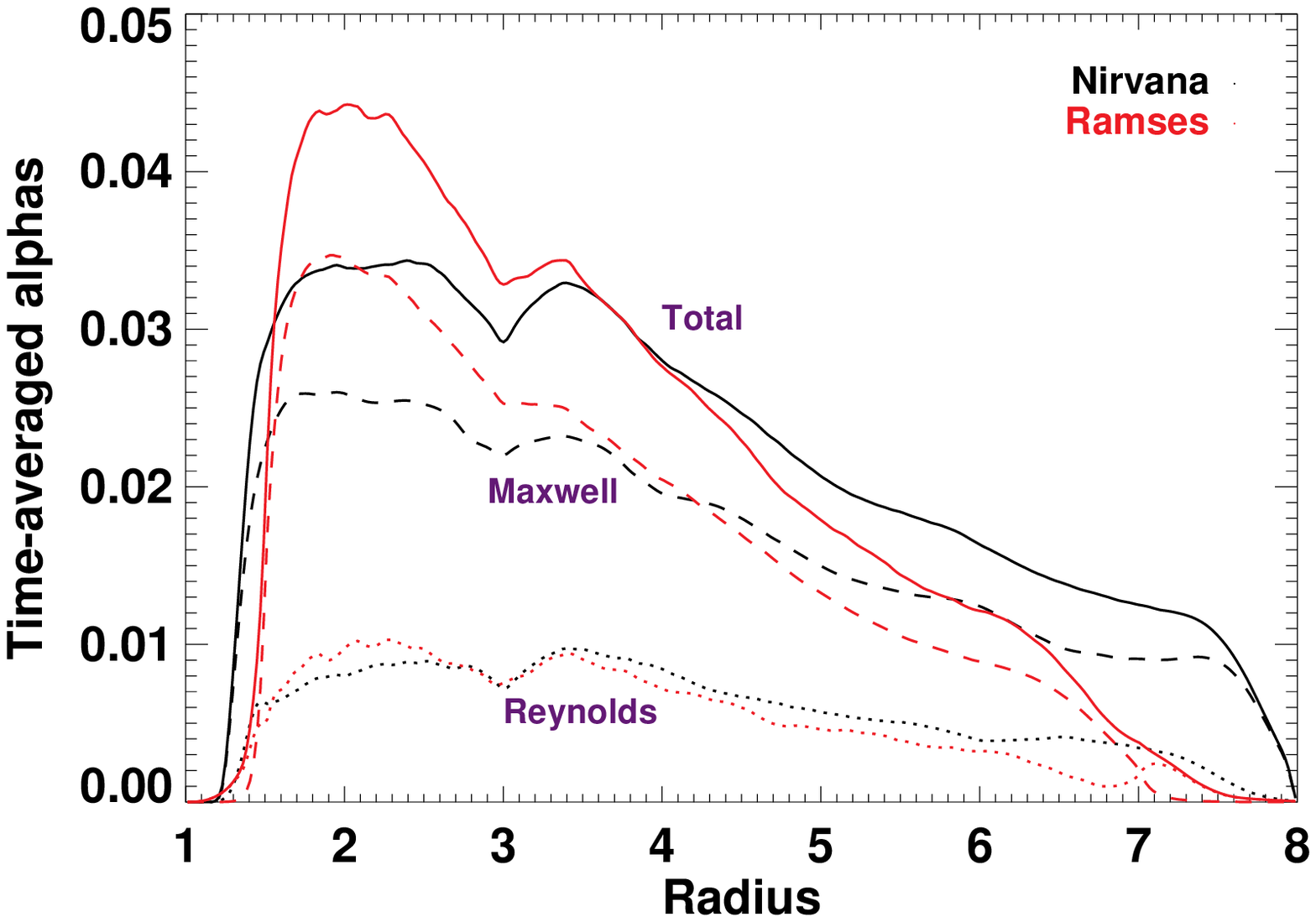}
   } 
   \caption{\label{fig:alpha_q1}Left: time evolution of the
     volume-averaged total alpha parameter, $\langle\alpha_{\rm
       tot}\rangle$, obtained with Model 1. Vertical dashed lines
     indicate the point in time when the planet is inserted in each
     simulation. Right: azimuthally- and vertically-averaged alpha
     parameters associated with the Reynolds stress (dotted curves),
     the Maxwell stress (dashed curves) and the total stress (solid
     curves) with the same model, time-averaged from 200 and 300
     orbits for the simulation with RAMSES, and from 250 to 350 orbits
     for the simulation with NIRVANA.}
\end{figure*}

The initial density and temperature profiles in Model 1 are the same
as those of the 2D simulations of Sect.~\ref{sec:resol}, where the
influence of the grid's azimuthal extent on the amplitude of the
corotation torque is examined. The time evolution of the
volume-averaged alpha parameter, associated with the sum of the
Reynolds and Maxwell stresses, is shown in the left panel of
Fig.~\ref{fig:alpha_q1}. It is denoted by $\langle\alpha_{\rm
  tot}\rangle$. The point in time from which the planet is inserted is
shown by a vertical dashed line for each simulation.  The time to
reach a turbulent saturated state is longer in the RAMSES run, since
it uses a larger $\beta$ value. In the saturated state,
$\langle\alpha_{\rm tot}\rangle \approx 0.02$ in both the NIRVANA and
RAMSES calculations. At first, this agreement might seem surprising
since the initial toroidal flux is different in the two simulations.
This flux is indeed conserved throughout the simulations, and it is
well known \citep[e.g.,][]{nelson05} that, in the saturated state of
the turbulence, $\langle\alpha_{\rm tot}\rangle$ scales with the
initial toroidal flux. This agreement is in fact due to the different
algorithm used by the two codes. Being less diffusive, Godunov codes
are known to give in general larger alpha values
\citep[e.g.,][]{Simon09}. In this study, we tuned the respective
initial toroidal fields in order to obtain the same saturated
transport properties.

\begin{figure}
 \resizebox{\hsize}{!}{\includegraphics{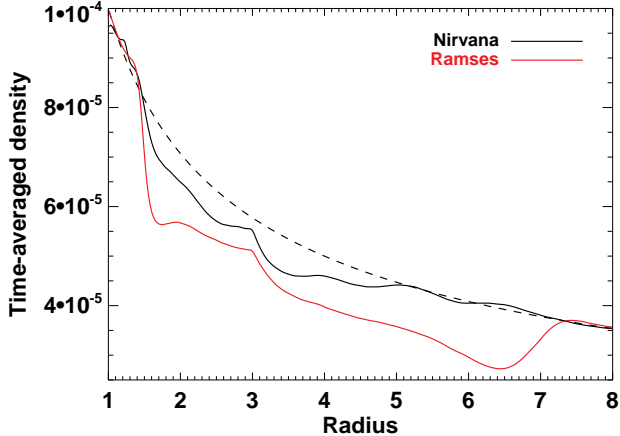}}
 \caption{\label{fig:rtadens_q1}Azimuthally- and time-averaged
   midplane density profile obtained in Model 1 with NIRVANA (black
   curve) and RAMSES (red curve). The time averaging is done between
   250 and 350 orbits for the NIRVANA run, and between 200 and 300
   orbits for the RAMSES run. The dashed curve displays the initial
   density profile common to both simulations.}
\end{figure}

The time-averaged radial profiles of the alpha parameters associated
with the Reynolds, Maxwell and total stresses (averaged over $\varphi$
and $z$) are displayed in the right panel of Fig.~\ref{fig:alpha_q1}.
They are respectively denoted by $\overline\alpha_{\rm Rey}$,
$\overline\alpha_{\rm Max}$ and $\overline\alpha_{\rm tot}$. The time
average is done between 200 and 300 orbits for the RAMSES run, and
between 250 and 350 orbits for the NIRVANA run. Note the slight
decrease in both $\overline\alpha_{\rm Rey}$ and $\overline\alpha_{\rm
  Max}$ near the planet's orbital radius, which accounts for the
slight increase in the disc density (or, equivalently, the disc
thermal pressure) at the same location, as shown in
Fig.~\ref{fig:rtadens_q1}. We checked that the time-averaged magnetic
stress is slightly increased inside the planet's Hill radius and at
the location of the planet's wake, where the magnetic field tends to
be compressed and ordered. This is reminiscent of the findings by
\cite{qmwmhd2} (see their figure 13), who however considered a planet
$\sim 15$ times more massive than ours.  At the planet location,
$\overline\alpha_{\rm tot} \sim 0.03$. This translates into an
averaged timescale for viscous diffusion across the horseshoe region
$\simeq x_{\rm s}^2 \,(\overline\alpha_{\rm tot} c_{\rm s} H)_{\rm
  p}^{-1} \approx 2.5 T_{\rm orb}$, which is similar to the horseshoe
U-turn timescale, $\tau_{\rm U-turn}\simeq 2 T_{\rm orb}$, see
Eq.~(\ref{eq:tauUturn}). We thus expect the corotation torque to adopt
a value close to that predicted by linear theory.

The time- and azimuthally-averaged disc midplane density is depicted
in Fig.~\ref{fig:rtadens_q1}.  The time-average is done over the same
time interval as for the total alpha parameter shown in the right
panel of Fig.~\ref{fig:alpha_q1}.  The mass-adding procedure described
in Sect.~\ref{sec:physmodel}, which relaxes the density profile
towards its initial value over 20 local orbital periods, implies that
we get a stationary density profile slightly reduced compared to the
initial one, depicted by a dashed curve in
Fig.~\ref{fig:rtadens_q1}. Note that the decrease in the disc mass is
larger in the RAMSES simulation at all radii. At the planet location,
the averaged-to-initial density ratio is $\approx 0.95$ and $\approx
0.88$ for the NIRVANA and RAMSES calculation, respectively.

\begin{figure*}
  \resizebox{\hsize}{!}{\includegraphics{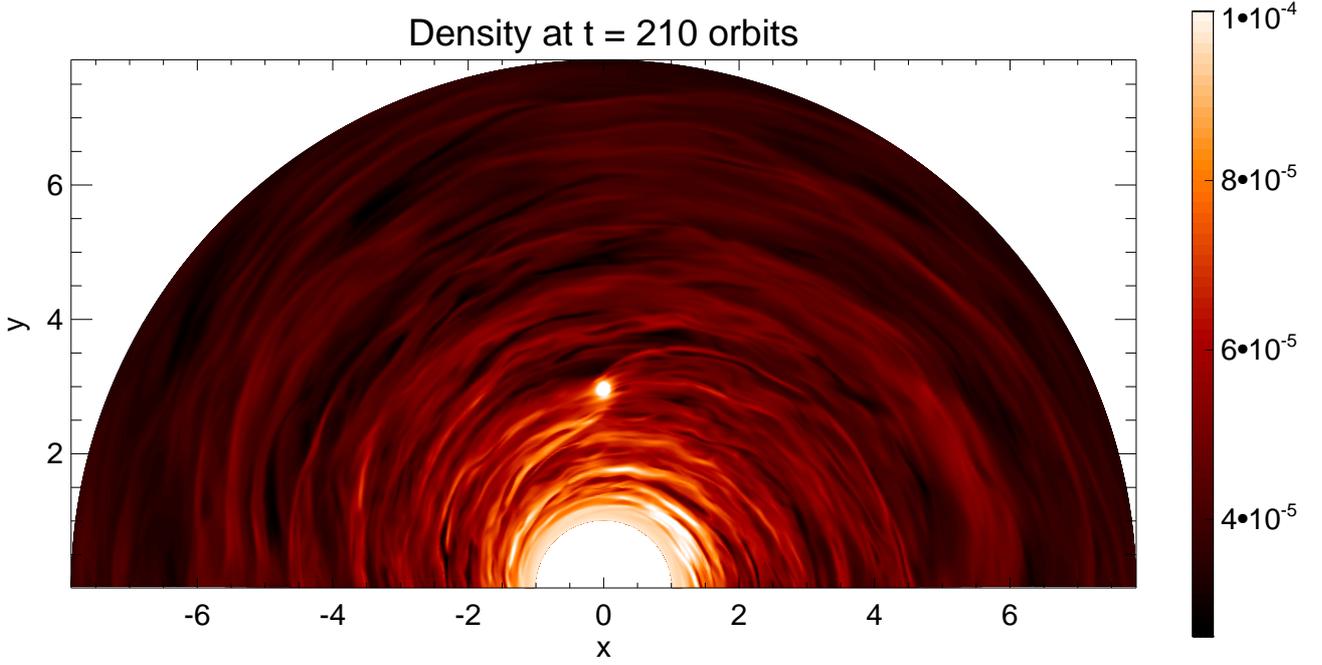}}
  \caption{\label{fig:niceplot}Density contours in the disc midplane
    shown at 210 orbits for the NIRVANA run with Model 1. The planet
    is located at $x=0$, $y=3$.}
\end{figure*}

\begin{figure*}
\centering\resizebox{\hsize}{!}
  {
    \includegraphics{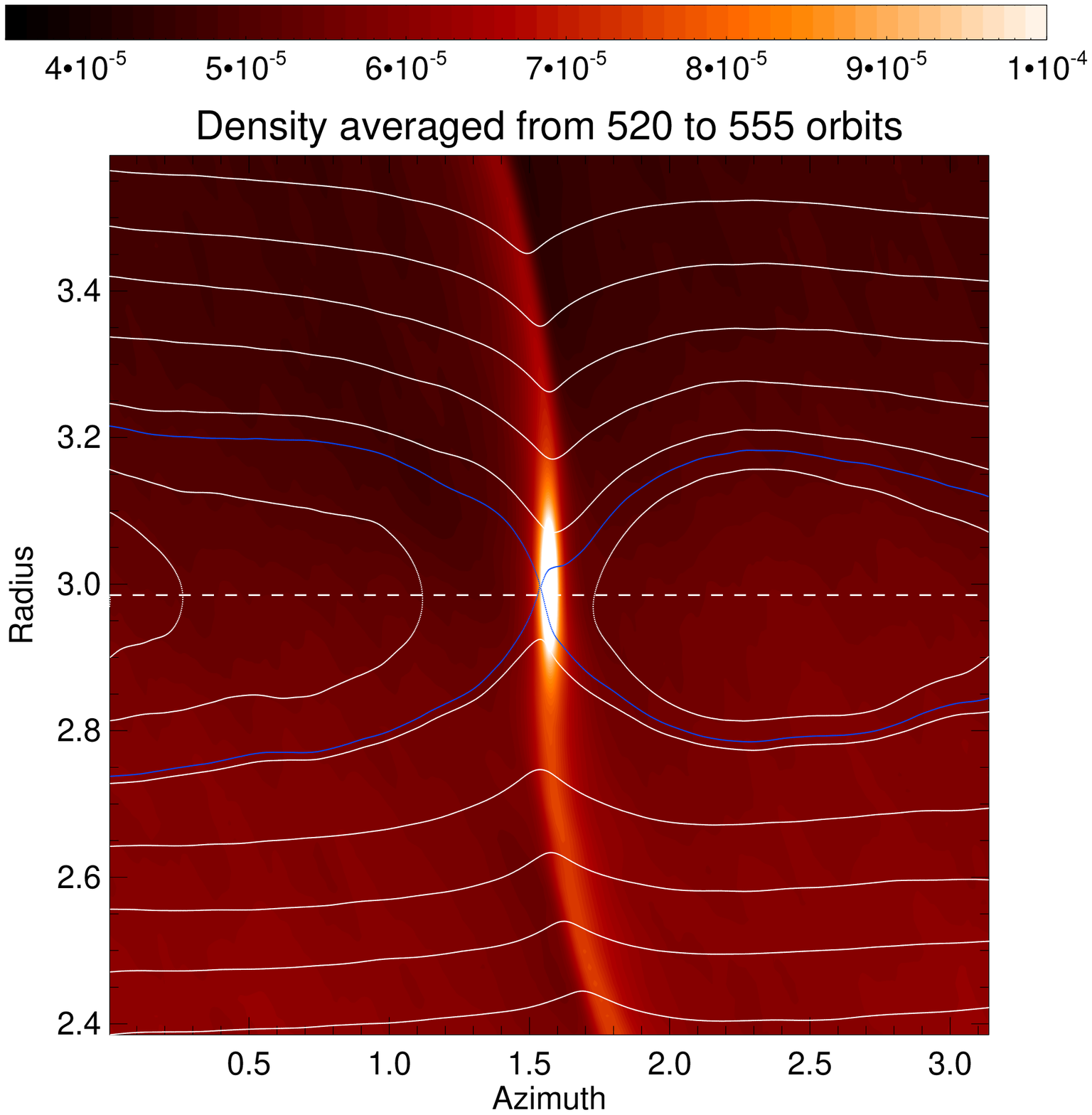}
    \includegraphics{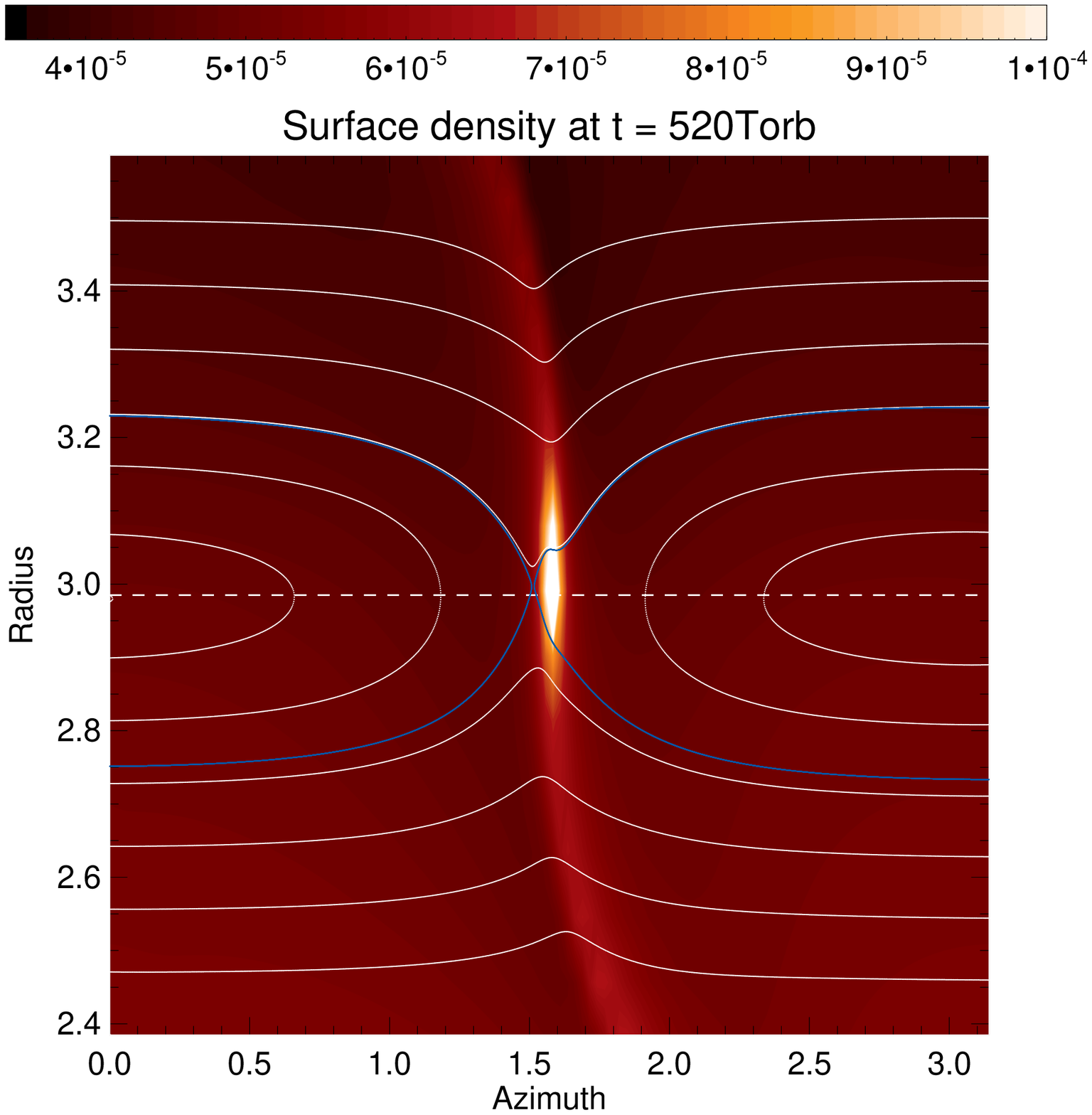}
   }
   \caption{\label{fig:strl}Left: contours of the midplane density
     obtained in Model 1 with the NIRVANA run, and time-averaged
     between 520 and 555 orbits. Streamlines in the disc midplane,
     time-averaged over the same time interval, are overplotted by
     solid curves. Right: gas surface density obtained at 520 orbits
     with disc Model 1 in the FARGO viscous run, with instantaneous
     streamlines overplotted. In both panels, the horizontal dashed
     line shows the planet's corotation radius, defined as the
     location $R_{\rm c}$ where $\Omega(R_{\rm c}) = \Omega_{\rm p}$.}
\end{figure*}

Density contours in the disc midplane obtained with the NIRVANA run
after 210 orbits are shown in Fig.~\ref{fig:niceplot}. The planet is
located at $x=0$, $y=3$, and the color scale has been adjusted such
that the maximum density contour is the maximum value of the initial
gas density. The spiral density waves generated by the planet are
hardly visible, their density contrast being very similar to that of
the turbulent fluctuations. Time-averaged density contours are
displayed in the left panel of Fig.~\ref{fig:strl}. The time averaging
is done over 35 orbits, starting 345 orbits after the restart time
(or, 520 orbits after the beginning of the simulation). This time
interval corresponds to $\sim 3$ horseshoe libration periods.  While
the grid's full azimuthal range is shown by the y-axis, the x-axis
displays a narrow range of radii about the planet's orbital radius.
The planet is located at $R_{\rm p}=3$ and $\varphi_{\rm p} = \pi/2$,
and its wake is clearly visible. Overplotted by solid lines are
streamlines in the disc midplane, evaluated from the time-averaged
values of the radial and azimuthal components of the gas velocity
field over the same time interval. Horseshoe streamlines are clearly
seen, which delimit a well-defined mean horseshoe region, encompassed
by mean circulating streamlines. The use of a quite long (35 orbits)
time interval for the time averaging is aimed at displaying smooth
averaged streamlines, which helps estimate the half-width of the
planet's mean horseshoe region, $\overline{x}_s$. As will be shown in
Sect.~\ref{sec:hs}, horseshoe streamlines are readily observed when
averaging over 5 orbits (see Fig.~\ref{fig:pass-scal}). The blue
streamlines depict the approximate location of the separatrices of the
mean horseshoe region. Its half-width at $\varphi - \varphi_{\rm p} =
1$ rad is $\overline{x}_s \approx 0.20$. For comparison, gas surface
density contours obtained with the FARGO viscous run are shown in the
right panel of Fig.~\ref{fig:strl} along with instantaneous
streamlines. We see that the width and shape of the mean horseshoe
region in the MHD turbulent simulation are very similar to their
counterpart in the laminar viscous simulation. In the viscous run, the
half-width of the horseshoe region is $x_{\rm s} \approx 0.23$ at
$\varphi - \varphi_{\rm p} = 1$ rad. This is in good agreement with
the value obtained from the expression given by \cite{pbck10} (their
Eq. 44),
\begin{equation}
  x_{\rm s} = 1.1R_{\rm p} \times \left(\frac{0.4}{\varepsilon/H_{\rm p}}\right)^{1/4} 
  \left( \frac{q}{h_{\rm p}} \right)^{1/2} \simeq 0.215.
\label{xs}
\end{equation}

\begin{figure*}
  \resizebox{0.5\hsize}{!}{\includegraphics{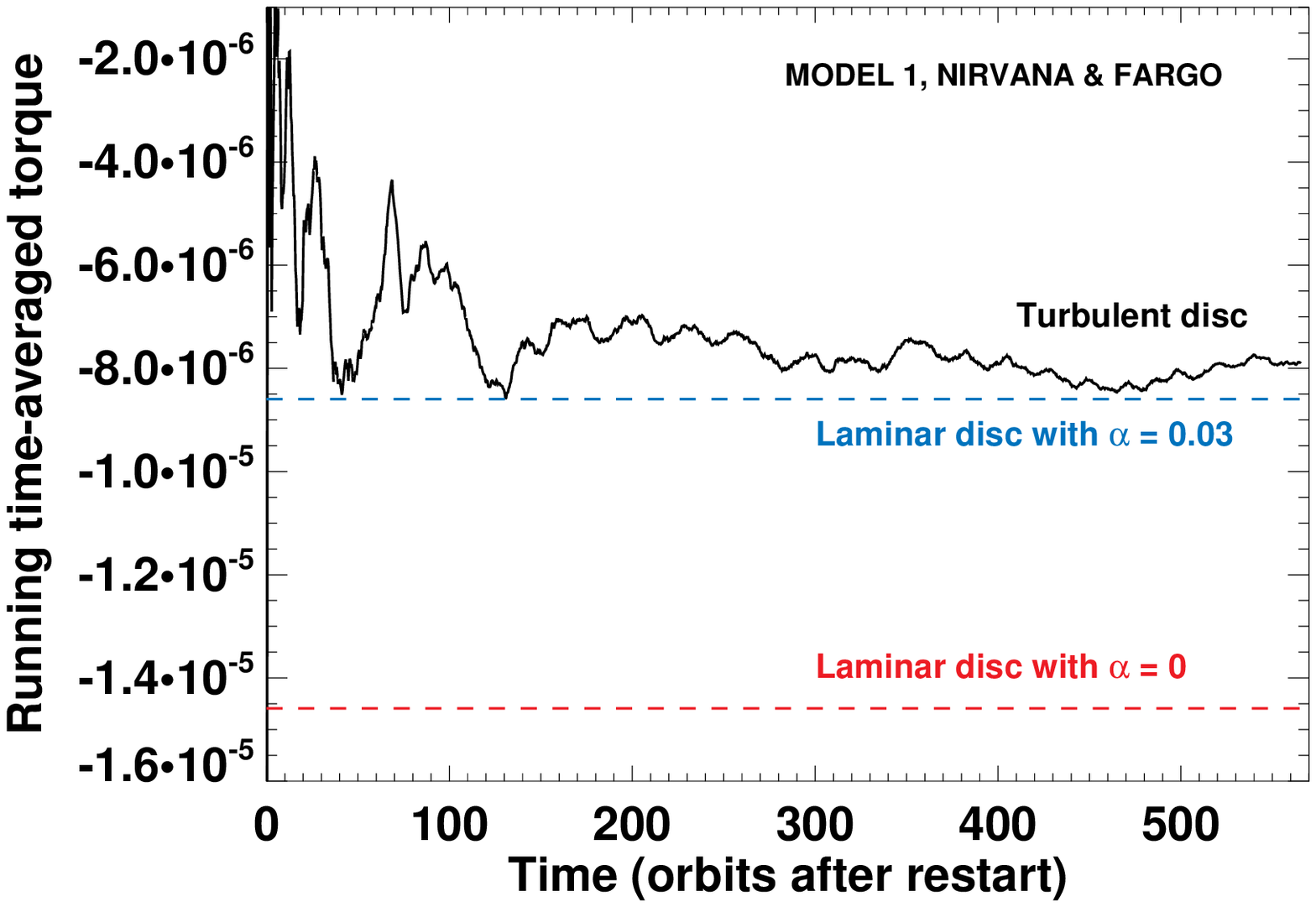}}
  \resizebox{0.5\hsize}{!}{\includegraphics{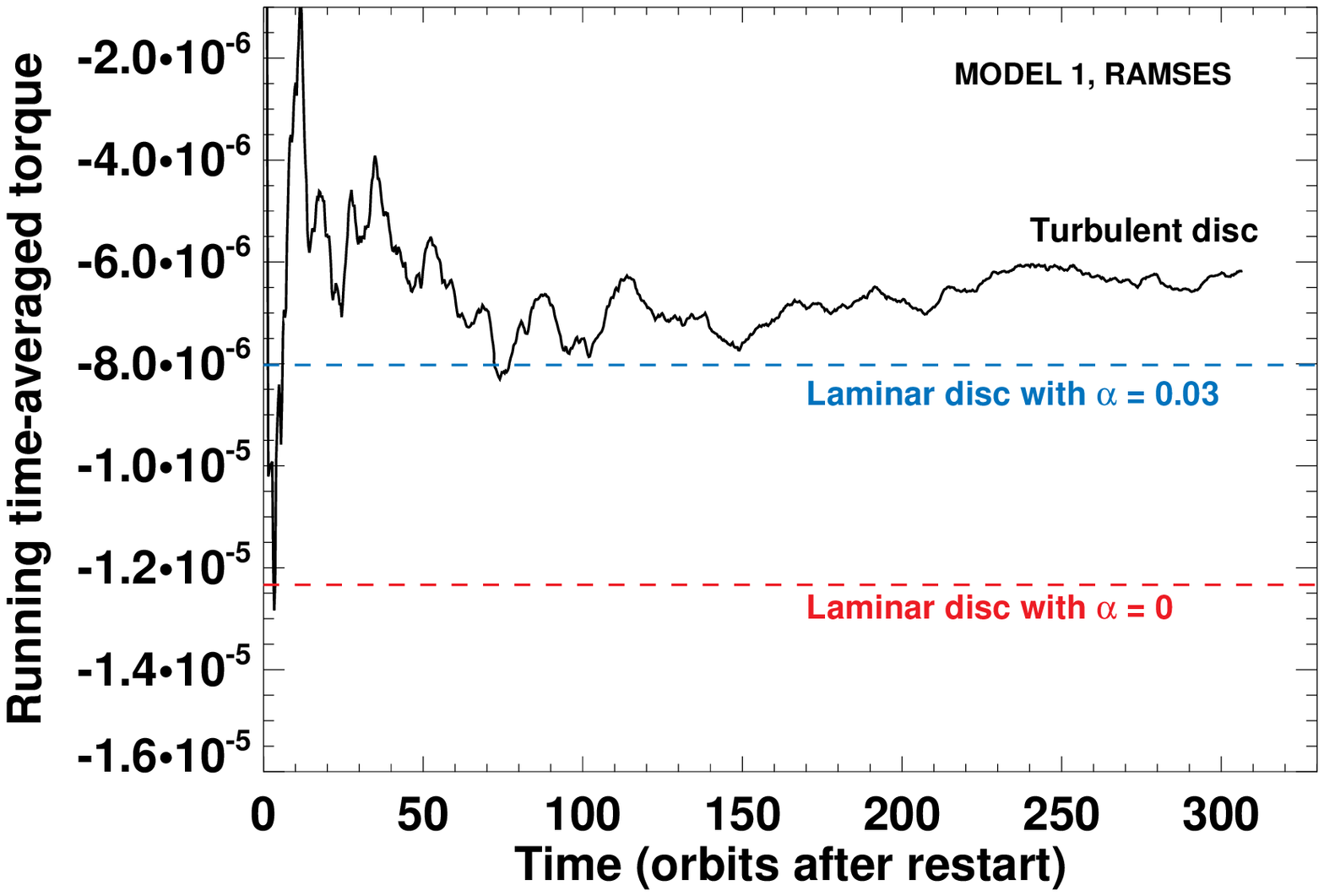}}
  \caption{\label{fig:rtatq_q1}Running time-averaged specific torque
    obtained with Model 1, for the NIRVANA run (left panel), and the
    RAMSES run (right panel). For comparison, the final value of the
    time-averaged torques obtained from 2D laminar disc models are
    overplotted, wherein the initial density profile is the
    time-averaged density profile obtained with the MHD run. Note that
    in the left panel, the turbulent run with NIRVANA is compared to
    laminar runs carried out with FARGO. Two series of laminar disc
    models have been performed, one with an inviscid disc
    ($\alpha_{\rm p}=0$), and one with a viscous disc (corresponding
    to $\alpha_{\rm p}=0.03$).}
 \end{figure*}
Fig.~\ref{fig:rtatq_q1} depicts the running time-averaged torque
(solid curves) obtained with NIRVANA (left panel) and RAMSES (right
panel).  We briefly recall that all torques shown in this paper refer
to the specific torque exerted by the disc on the planet, and that the
content of the planet's Hill radius is excluded in the torque
calculation. In both runs, the torque reaches a stationary value after
typically 100 to 200 orbits. Although not shown here, this timescale
is consistent with the amplitude of the stochastic torque standard
deviation being similar to that of the type I migration torque in an
equivalent laminar disc model \citep[e.g.,][]{np2004, bl10}. This
similarity can be qualitatively expected from the similar density
contrasts of the turbulent fluctuations and of the planet wake,
illustrated in Fig.~\ref{fig:niceplot}. We comment that the stationary
running time-averaged torque obtained with both codes are in very good
agreement, that of the RAMSES run being 10 to $20\%$ smaller than that
of the NIRVANA run. This slight difference may be (at least partly)
attributed to the time-averaged density profile being slightly smaller
in the RAMSES run (see Fig.~\ref{fig:rtadens_q1}).

How do the averaged torques of our MHD turbulent simulations compare
with the predictions of laminar disc models? To address this question,
we have performed two series of 2D laminar disc calculations using
FARGO and RAMSES. In one series, the disc is inviscid.  In the other
series, a viscous disc model is used with a prescribed kinematic
viscosity $\nu$. In the latter case, the FARGO run uses a fixed radial
profile for $\alpha$ that matches the time-averaged total alpha
profile obtained in the MHD run with NIRVANA (black curve in the right
panel of Fig.~\ref{fig:alpha_q1}). The RAMSES run uses a constant
kinematic viscosity, chosen such that $\alpha_{\rm p}$ equals the
averaged alpha value obtained in the RAMSES MHD run at the planet
location ($\approx 0.03$). In both series of laminar disc models, the
initial density profile is the time-averaged density profile of the
corresponding MHD simulations (shown in Fig.~\ref{fig:rtadens_q1}). We
have found that the steady-state density profiles in the viscous
simulations are reduced by about $10\%$ with respect to their initial
profile. At the end of the inviscid simulations (400 orbits), a tiny
gap has opened about the planet's orbital radius, with a density
contrast $\simeq 20\%$.

The final value of the running time-averaged torque obtained in both
series of laminar calculations is overplotted in
Fig.~\ref{fig:rtatq_q1} by dashed lines. The 2D laminar torques have
been multiplied by the scaling factor $f=0.6$ to account for the
vertical extent of the 3D simulations (where $z$ ranges from -0.3 to
0.3). The torques obtained in the turbulent and viscous disc models
are in good agreement, their relative difference being less than
$10\%$ for the NIRVANA and FARGO runs, and less than $20\%$ for the
RAMSES runs. Taking into account that the stationary density profile
of the viscous runs is $\simeq 10\%$ smaller than the time-averaged
density profile of the turbulent runs, the actual relative difference
between the viscous and turbulent torques ranges from $20\%$ to
$30\%$. In the series of inviscid runs, the torque is more negative,
as expected, since the (positive) corotation torque saturates after a
few libration timescales, leaving only the differential Lindblad
torque.
  
It is instructive to compare our results with the predictions of the
analytical torque expressions of \cite{pbck10}, derived for 2D viscous
disc models with power-law density profiles. At the end of the
inviscid simulations, although a small gap has opened, the density
profile can still be approximated as a power-law profile. For
instance, in the FARGO inviscid run, it is fairly well approximated by
$\Sigma_{\rm p} (R/R_{\rm p})^{-1/2}$ with $\Sigma_{\rm p} \simeq 5.2
\times 10^{-5}$. The torque expression in Eq.~(14) of \cite{pbck10}
predicts a specific differential Lindblad torque $\Gamma_{\rm L}
\simeq -6.8 (M_{\rm p} / M_{\star})\,h_{\rm p}^{-2}\,\Sigma_{\rm
  p}\,R^4_{\rm p}\,\Omega^2_{\rm p}$.  Multiplying this torque
expression by the aforementioned scaling factor $f$ gives $\Gamma_{\rm
  L} \approx -1.9 \times 10^{-5}$ for the density profile of the FARGO
inviscid run. This predicted value compares decently with our
simulation result ($\approx -1.45\times 10^{-5}$, see left panel of
Fig.~\ref{fig:rtatq_q1}). Given the large viscosity parameter in our
turbulent model, we expect that the corotation torque is close to its
linear value \citep{pp09a}.  Similarly, using the parameters of the
FARGO viscous run, \cite{pbck10} predict a linear corotation torque
$\approx 5.6\times 10^{-6}$ (see their Eq. 16). This is in good
agreement with our findings, where the corotation torque (estimated as
the torque of the viscous model substracted from the torque of the
inviscid model) amounts to $\approx 6.5\times 10^{-6}$.
  
\begin{figure*}
  \includegraphics[width=0.5\hsize]{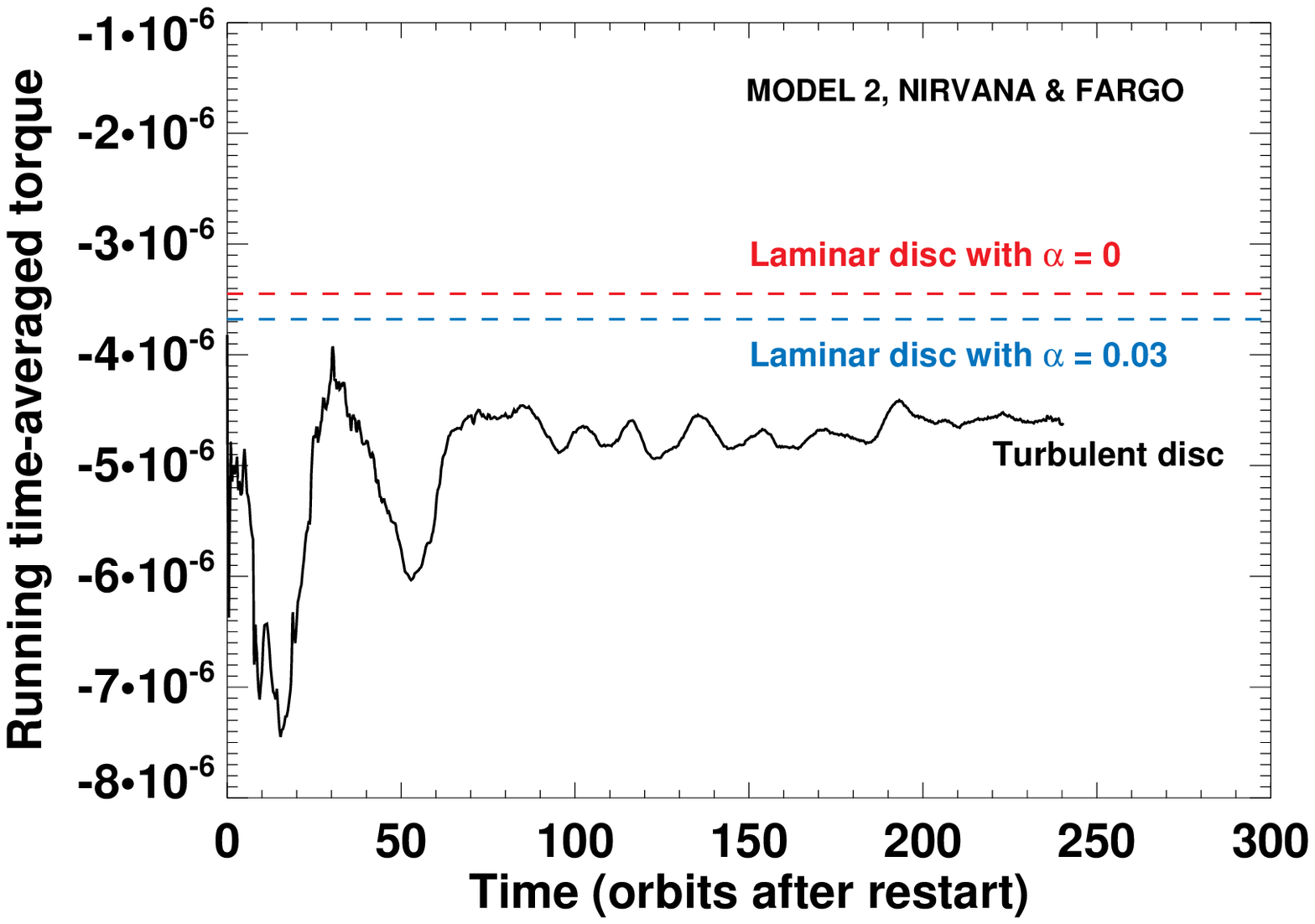}
  \includegraphics[width=0.5\hsize]{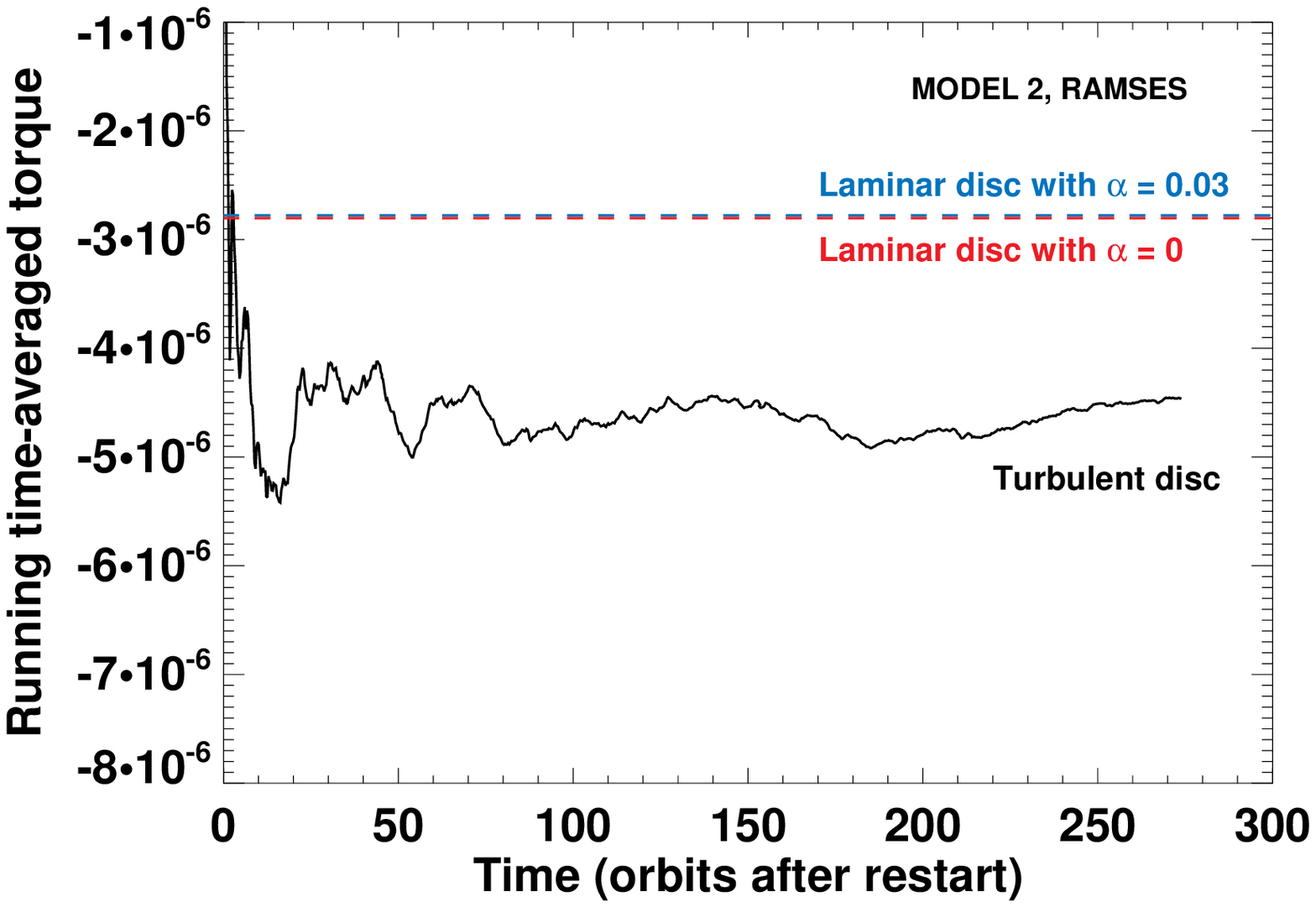}
  \caption{\label{fig:rtatq_q0}Running time-averaged specific torque
    obtained in Model 2 with the NIRVANA run (left panel) and the
    RAMSES run (right panel). As in Fig.~\ref{fig:rtatq_q1}, the
    results of 2D inviscid and viscous disc models are overplotted,
    wherein the initial density profile is the time-averaged density
    profile of the corresponding MHD run.}
 \end{figure*}
 The results of this section brings more support to the existence of
 an unsaturated corotation torque in weakly magnetized turbulent
 discs.  We find a decent agreement between the running time-averaged
 tidal torques in this particular MHD turbulent disc model, and in a
 viscous non-magnetic disc model with similar viscosity parameter at
 the planet location. The properties (shape, width) of the mean
 horseshoe region compare decently with those of the non-magnetic
 case. We cannot conclude, however, that the differential Lindblad
 torque and the corotation torque take similar values in both types of
 models. It is possible that MHD turbulence alters both the
 differential Lindblad torque and the corotation torque, in such a way
 that the total tidal torque is hardly changed. The results of Model
 2, where laminar disc models predict no corotation torque, will help
 us evaluate to what extent MHD turbulence alters the differential
 Lindblad torque, and also whether or not the inclusion of magnetic
 effects can introduce an additional torque.

 \subsection{Model 2: initial density $\propto R^{-3/2}$ and uniform
   temperature}
\label{sec:model2}
The results of Model 1, presented in Sect.~\ref{sec:model1}, strongly
indicate the existence of an unsaturated corotation torque in weakly
magnetized turbulent discs. The tidal torque compares decently with
the value expected in laminar viscous disc models. In this section, we
consider a disc model with a constant temperature, and an initial
density profile decreasing as $R^{-3/2}$. We refer to this model as
Model 2.  Laminar non-magnetic disc models predict no corotation
torque with such disc profiles, and that the tidal torque therefore
coincides with the differential Lindblad torque. For the alpha values
of our turbulent disc models (a few percent), the differential
Lindblad torque is expected to have no significant dependence with the
disc viscosity \citep[e.g.,][]{Muto09}.

As in Model 1, we have carried out 2D inviscid and viscous simulations
using the FARGO and RAMSES codes. Their initial density profile is the
time-averaged profile obtained in the corresponding 3D MHD runs (that
of the NIRVANA run is used as initial condition for the 2D run with
FARGO). The FARGO viscous run imposes a radial profile of alpha
viscosity equal to the radial profile $\overline{\alpha}_{\rm tot}$ of
the NIRVANA run. In the RAMSES viscous run, a constant kinematic
viscosity is used, such that the corresponding value of $\alpha_{\rm
  p}=0.03$, the value of $\langle\alpha_{\rm p}\rangle$ in the RAMSES
MHD simulation.

The time evolution of the running time-averaged torque is depicted in
Fig.~\ref{fig:rtatq_q0}. Results with NIRVANA and FARGO are displayed
in the left panel, those with RAMSES in the right panel. As in
Fig.~\ref{fig:rtatq_q1}, the torques of the 2D runs are multiplied by
the scaling factor $f=0.6$, to compare them with the torques of the 3D
runs. The stationary running time-averaged torques of both MHD
simulations are in close agreement. This is maybe partly fortuitous,
given that the corresponding time-averaged density profiles differ by
about $25\%$ (not shown here, but analogous to the different behaviors
obtained in Model 1, and shown in Fig.~\ref{fig:rtadens_q1}). This
$\simeq 25\%$ relative difference in density, however, is consistent
with the relative difference between the torques in the laminar
viscous simulations. The tiny torque difference between the inviscid
and viscous runs with FARGO is most probably due to the opening of a
shallow gap structure about the planet location. Gap formation also
occurs in the RAMSES run, but in this particular case, it is found to
have a negligible influence on the stationary running time-averaged
torque.

For the parameters of Model 2, the differential Lindblad torque
expression of \cite{pbck10} (see their Eq. 14) reads $\Gamma_{\rm L}
\simeq -3.9 (M_{\rm p} / M_{\star})\,h_{\rm p}^{-2}\,\Sigma_{\rm
  p}\,R^4_{\rm p}\,\Omega^2_{\rm p}$.  The time-averaged density
profiles of our turbulent simulations can be approximated as
$\Sigma_{\rm p} (R/R_{\rm p})^{-3/2}$ near the planet location, with
$\Sigma_{\rm p} \simeq 1.9\times 10^{-5}$ and $1.6\times 10^{-5}$ for
the NIRVANA and RAMSES runs, respectively. The value of $\Gamma_{\rm
  L}$ predicted by laminar disc models then amounts to about
$-3.9\times 10^{-6}$ and $-3.3\times 10^{-6}$ for the NIRVANA and
RAMSES runs, respectively. These predicted values agree with our
numerical results to within 10$\%$ for the NIRVANA run, and 30$\%$ for
the RAMSES run.

The running time-averaged torque of the NIRVANA turbulent run is in
reasonable agreement with the torque of the FARGO viscous run, their
relative difference being about $30\%$. The agreement is less good for
the RAMSES run, where the relative difference is $\sim 60\%$.
Interestingly, the averaged torque measured in both MHD simulations is
more negative than in the laminar viscous runs, while the opposite
trend is observed in Model 1. This comparison suggests that either the
differential Lindblad torque is slightly increased, or that there
exists an additional torque in magnetized turbulent discs. More
insight into this hypothesis is given in Sect.~\ref{sec:tqdistri}.

\section{Discussion}
\label{sec:discu}

\subsection{Horseshoe dynamics}
\label{sec:hs} 
Although the results presented in previous sections highlight the
action of unsaturated corotation torques, we have not so far
demonstrated the existence of horseshoe dynamics in our simulations
without time-averaging over long timescales. A key issue is whether
the horseshoe trajectories exist and can be detected easily in the
presence of MHD turbulence and the associated turbulent velocity
fluctuations. A related issue is the question of how rapidly the
turbulence causes diffusion across the horseshoe region. This is a key
factor in determining the magnitude of the corotation torque. As
discussed in Sect.~\ref{sec:resol}, the corotation torque in laminar
viscous disc models should have its maximal non-linear value when the
diffusion timescale across the horseshoe region, $\tau_{\rm visc}$, is
shorter than the horseshoe libration time, $\tau_{\rm lib}$, and is
larger than the horseshoe U-turn time $\tau_{\rm U-turn} \simeq h_{\rm
  p} \tau_{\rm lib}(\Delta\varphi=2\pi)$. If the diffusion time
$\tau_{\rm visc} \le \tau_{\rm U-turn}$, however, the corotation
torque will approach the lower value expected from linear theory.

\begin{figure*}
\centering\resizebox{\hsize}{!}{\includegraphics{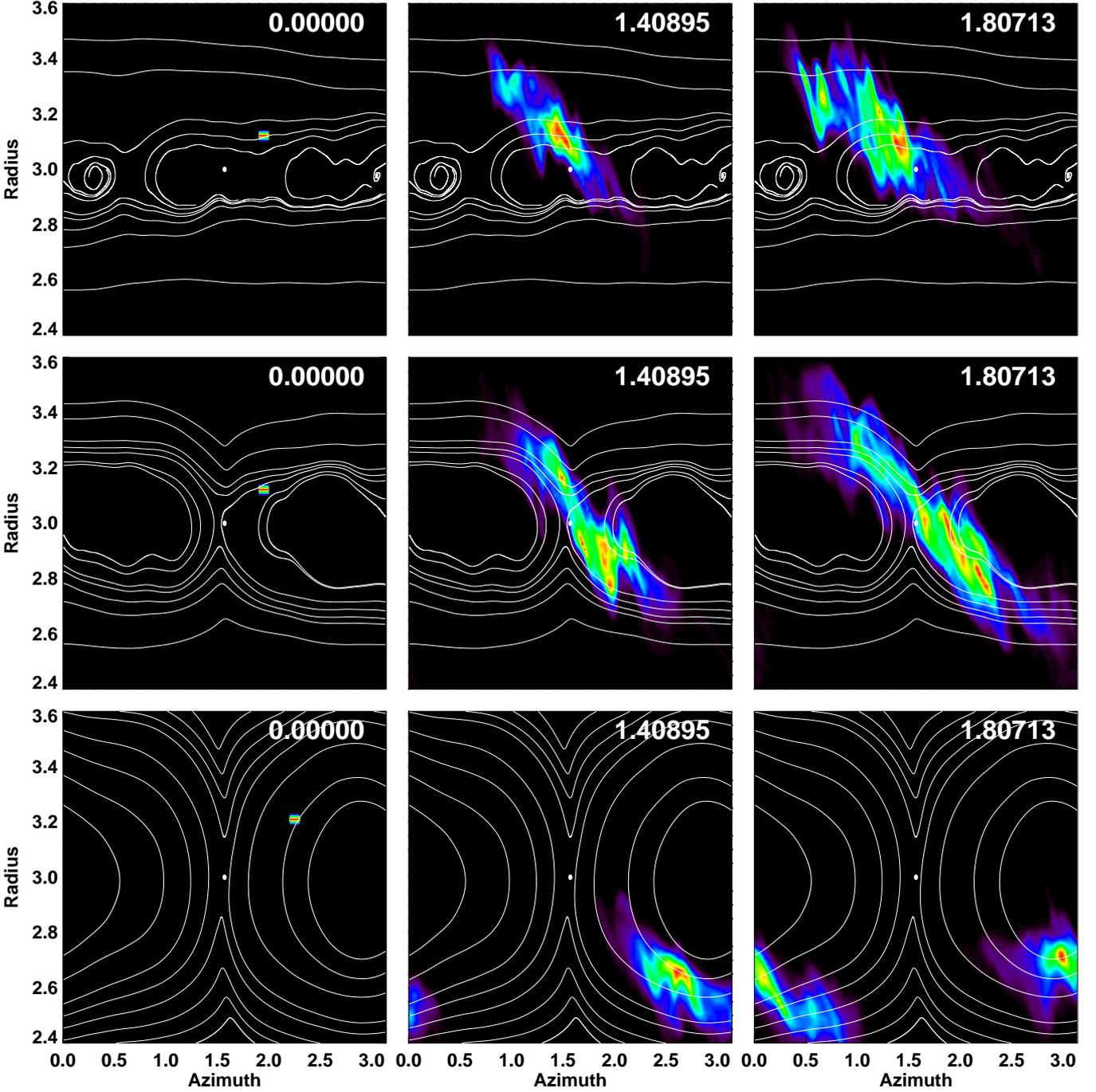}}
\caption{\label{fig:pass-scal} Evolution of the concentration of the
  passive contaminant, $C(R,\varphi)$, defined in the text, with
  superimposed averaged streamlines in the midplane for disc models
  with parameters equal to those of Model 1. The contour levels are
  defined for each panel separately to highlight the region occupied
  by the contaminant at each time.  The upper three panels are for a
  disc with no embedded planet.  The middle three panels have an
  embedded planet with $M_{\rm p}=3 \times 10^{-4}M_{\star}$. The
  lower panels are for an embedded planet with $M_{\rm
    p}=10^{-3}M_{\star}$. Times are given in units of the planet
  orbital period. Results are obtained with NIRVANA.}
\end{figure*}
In order to examine the diffusion rate across the horseshoe region, we
have performed simulations in which a small patch of the disc in the
horseshoe region is polluted initially with a passive contaminant that
is advected with the flow. We define a function $\psi(R, \varphi)
=C(R, \varphi) \, \rho$, where $C(R, \varphi)$ is the concentration of
the contaminant, and evolve the equation
\begin{equation}
  \frac{\partial \psi}{\partial t} + \nabla \cdot (\psi {\bf v}) = 0.
\label{eqn:scalar}
\end{equation}
We have carried out three simulations to examine the effects of
turbulent diffusion. The disc model in each run is the same as Model 1
(see Sect.~\ref{sec:model1}).  In the first run, no planet was
included in the simulation.  In the second run, a planet with $M_{\rm
  p} = 3 \times 10^{-4}M_{\star}$ was included, and the system was
relaxed for approximately 40 planet orbits prior to the contaminant
being introduced. In the third simulation, a Jovian mass planet with
$M_{\rm p} = 10^{-3}M_{\star}$ was included, and a similar process of
disc relaxation was undertaken to ensure that horseshoe trajectories
were fully established prior to the release of the contaminant.

The results of these calculations are shown in
Fig.~\ref{fig:pass-scal}, which displays contours of the contaminant
concentration in the disc midplane, at three different times.  The top
panels show the run with no planet included. The initial patch of
contaminant is located at $R=3.08$, between $1.9 \le \varphi \le 2.0$,
and at all heights in the disc. Moving from left to right through the
panels we see that the turbulence causes quite rapid diffusion in the
radial direction, with the contaminant diffusing over a radial
distance $\simeq 0.2$ during a time of 1.41 orbits.  The Keplerian
shear combines with the radial diffusion to cause the contaminant to
be stretched along the azimuthal direction through
advection. Nonetheless, we are able to observe that on average the
initial patch of contaminant continues along its original trajectory,
shearing past the nominal position of the planet that is indicated by
the filled white circle.  We have superimposed streamlines calculated
from the time-average of the velocity field in the disc midplane over
5 orbital periods measured at $R=3$ (the orbital radius of the planet,
if present).  As we are working in the rotating frame centred on this
radius, these show the Keplerian shear for disc regions that are
interior and exterior to $R=3$, but it is also clear that irregular
motions associated with the turbulence have not been completely
averaged out over this time.

The middle panels of Fig.~\ref{fig:pass-scal} show the run with
$M_{\rm p}=3 \times 10^{-4}M_{\star}$. It is immediately obvious that
the time-averaged velocity field displays the existence of horseshoe
streamlines. In general, we find that time averaging is required for
temporal windows of between $3-5$ orbits before the horseshoe
streamlines become clearly defined in Model 1, for this planet mass.
The half-width of the horseshoe region is approximately equal to 0.2,
in agreement with the previous determination from the left panel of
Fig.~\ref{fig:strl}, where streamlines are time-averaged over 35
planet orbits.

We see that the evolution of the passive scalar is rather different
from that observed in the no-planet case, and although there is
significant radial diffusion we also see clear evidence of the fluid
following the horseshoe streamlines. In particular, the presence of
significant contaminant near the downstream inward separatrix (that
is, the region bounded by $2.8 \le R \le 3$ and $\pi/2 \le \varphi \le
2$) in the second and third panels demonstrates that the fluid has
undergone a U-turn after approaching the planet. Similarly, at $t \sim
1.81 T_{\rm orb}$, the concentration in the region bounded by $3.0 \le
R \le 3.4$ and $0.5 \le \varphi \le \pi/2$ is smaller than in the no
planet case. For our disc model and planet parameters, $\tau_{\rm lib}
\simeq 10 T_{\rm orb}$, $\tau_{\rm U-turn} \simeq 2 T_{\rm orb}$, and
Fig.~\ref{fig:pass-scal} indicates that $\tau_{\rm visc} \gtrsim
\tau_{\rm U-turn}$. Note that turbulent diffusion of a passive scalar
is powered by the Reynolds stress only. However, the turbulent
diffusion of vortensity, which should control the magnitude of the
corotation torque by analogy with viscous disc models, depends \emph{a
  priori} both on the Reynolds and the Maxwell stresses.  The
timescale for turbulent diffusion of vortensity should thus be shorter
than for the contaminant, and we expect the corotation torque to have
a value smaller than its fully unsaturated non-linear value, in
agreement with our earlier discussion of Model 1.

The lower panels of Fig.~\ref{fig:pass-scal} show results for the run
with $M_{\rm p}=10^{-3}M_{\star}$, and we include this simulation in
our discussion only for illustration purposes. The horseshoe
streamlines are now much more clearly defined, and the trajectory of
the passive contaminant follows these streamlines even more clearly
than in the $M_{\rm p} = 3\times 10^{-4}M_{\star}$ run.  Here,
advection of the passive contaminant around the horseshoe streamlines
occurs more rapidly than radial diffusion. For the time scales covered
by the panels in Fig.~\ref{fig:pass-scal}, the contaminant therefore
remains more or less contained within the horseshoe region.
\begin{figure*}
\centering\resizebox{\hsize}{!}
  {
    \includegraphics{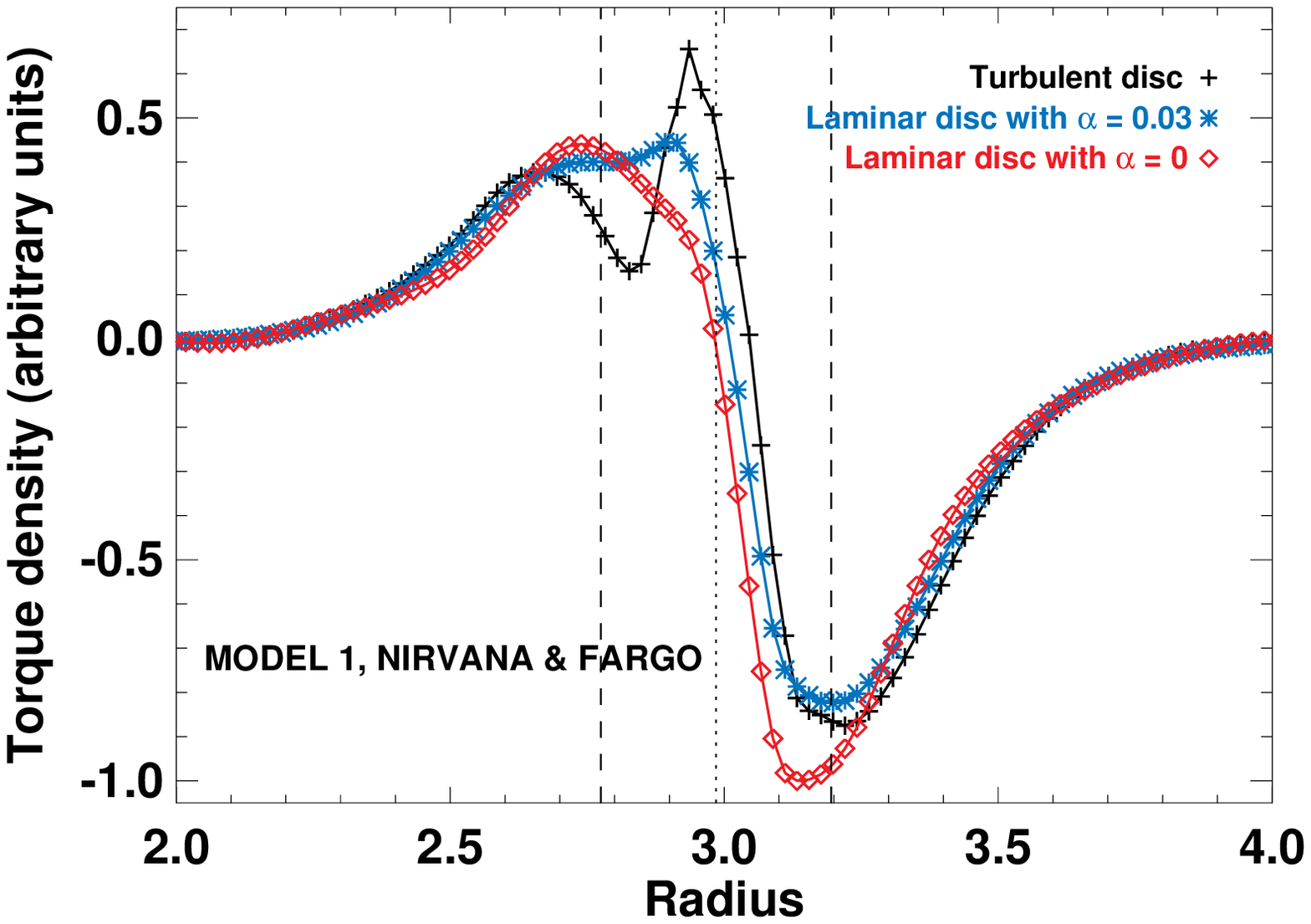}
    \includegraphics{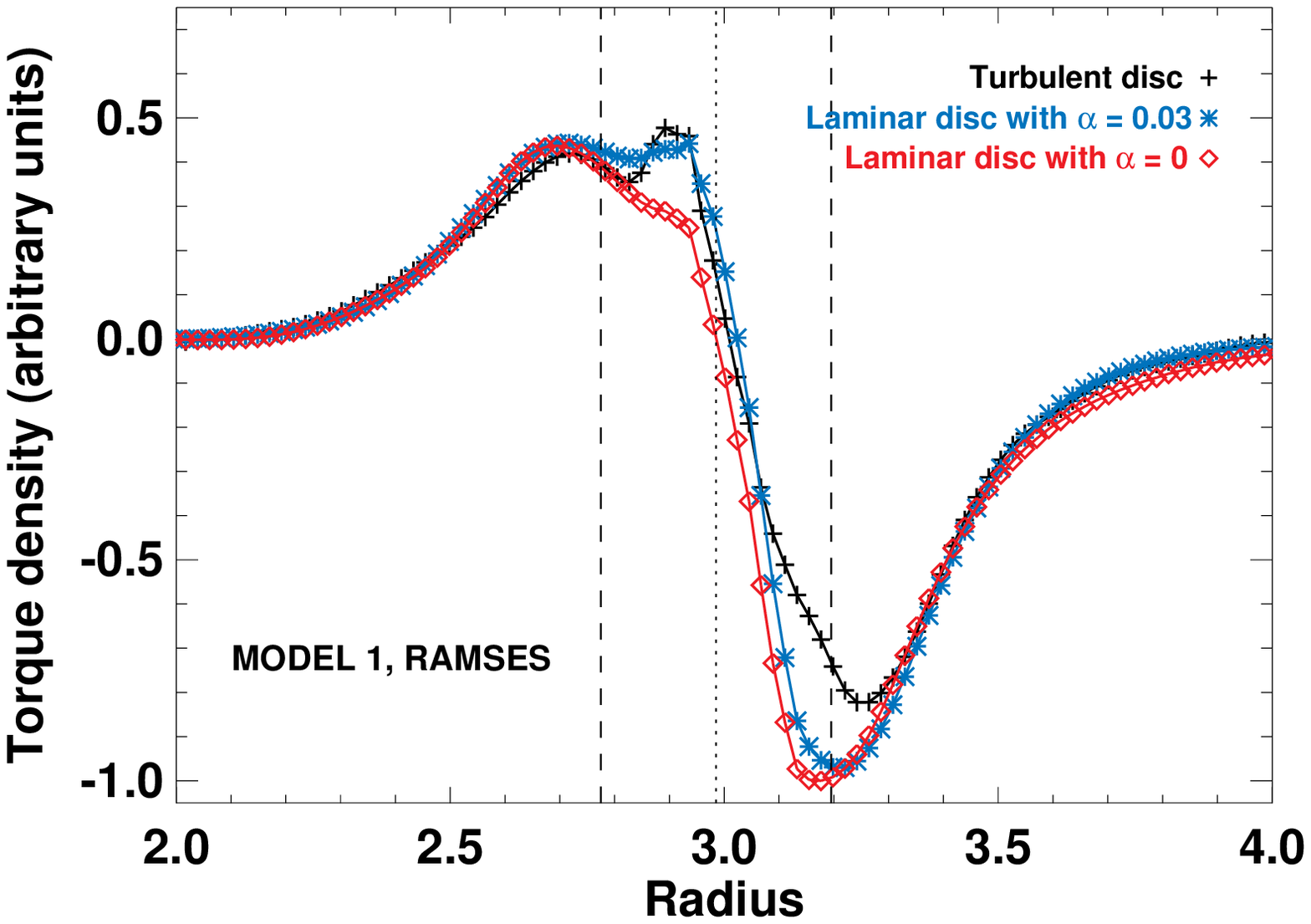}
   }
    \centering\resizebox{\hsize}{!}
  {
    \includegraphics{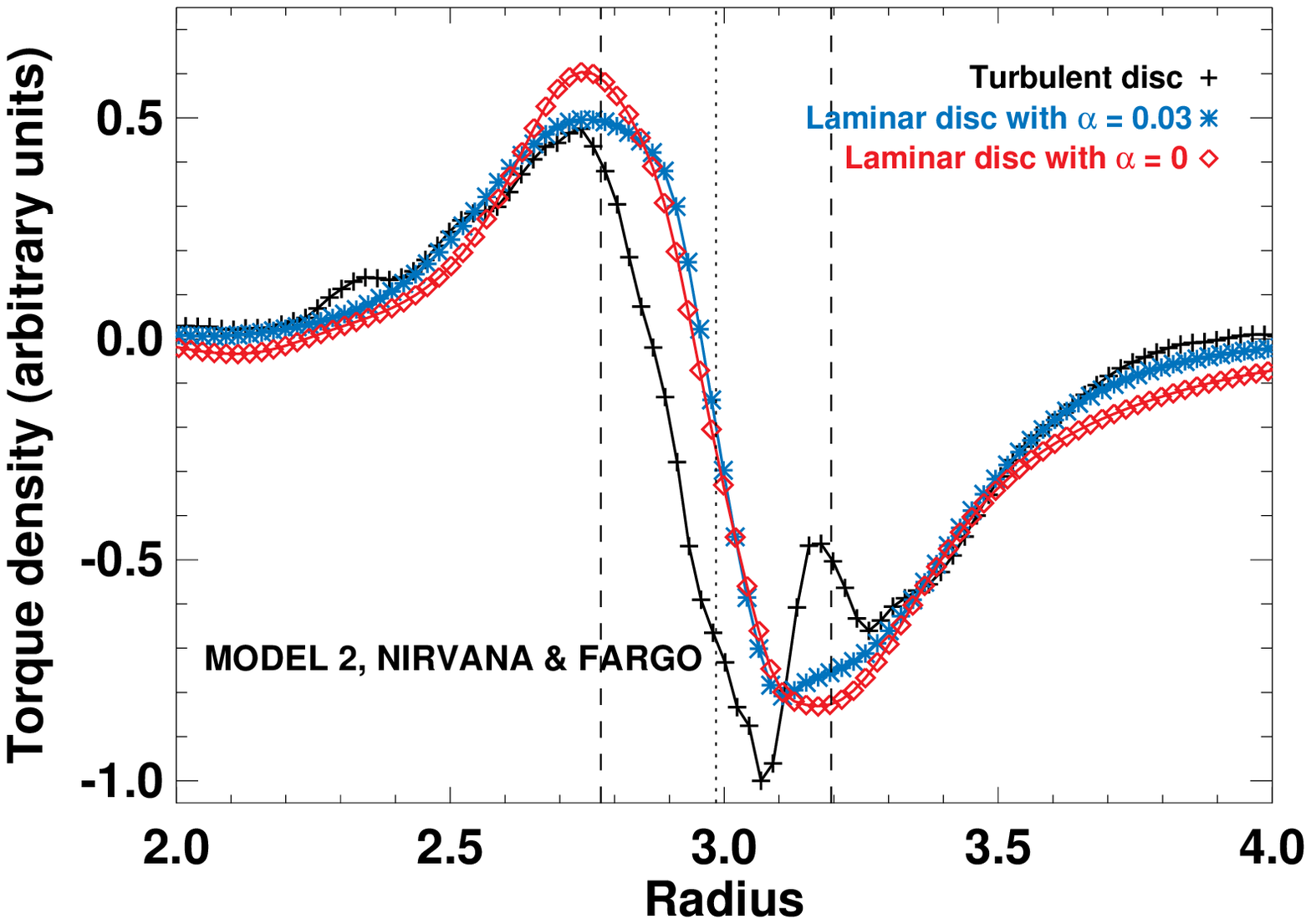}
    \includegraphics{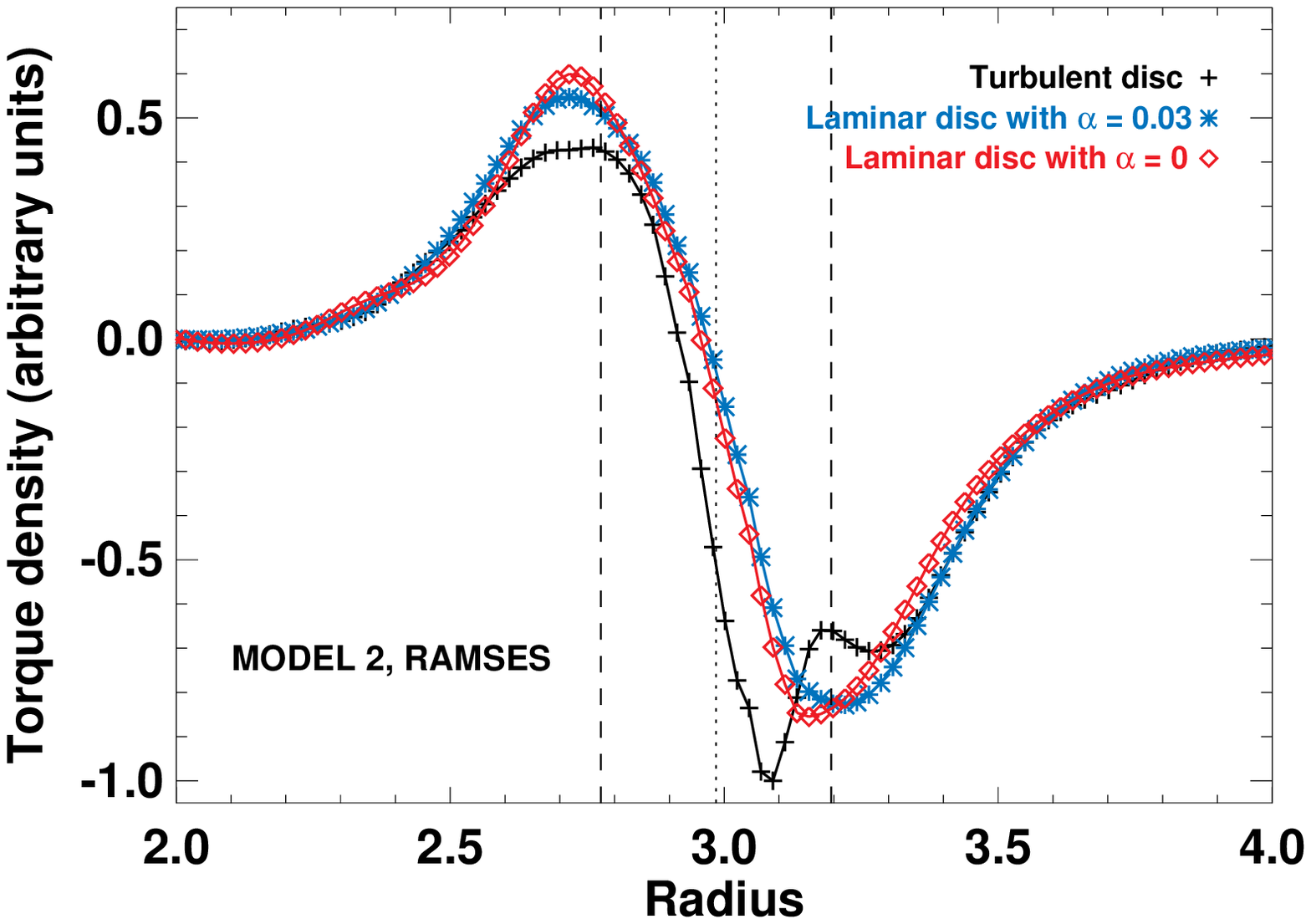}
  }
  \caption{\label{fig:tqprof}Time average of the torque density
    distribution, obtained with Model 1 (top row, plus signs) and
    Model 2 (bottom row, plus signs). For comparison, the torque
    density distribution of the inviscid and viscous non-magnetic disc
    models are overplotted (diamond and star symbols, respectively).
    The torque profiles of the turbulent runs have been time-averaged
    over 100 orbits. In each panel, torque density profiles are
    divided by the maximum of the (absolute value of the) inviscid,
    viscous, and turbulent torque profiles. In x-axis, orbital
    separation is depicted from $R=2$ to $R=4$, which corresponds to a
    radial extent $\simeq \pm 3H_{\rm p}$ about the planet
    location. In each panel, the dashed lines show the approximate
    location of the separatrices of the planet's horseshoe region, and
    the dotted line that of the planet's corotation radius.}
 \end{figure*}

 For the disc and planet parameters of our model, we find clear
 evidence of the existence of horseshoe dynamics in weakly magnetized
 turbulent discs, in the sense that mean horseshoe streamlines are
 exhibited along which fluid elements perform horseshoe U-turns
 despite substantial turbulent diffusion. We comment that we have not
 shown however whether it is possible or not to get a fully
 unsaturated horseshoe drag in such turbulent configurations. Also,
 the presence of horseshoe dynamics in turbulent discs can be
 attributed to the averaged turbulent velocity fluctuations being of
 smaller amplitude than the U-turn drift rate. It is unclear whether
 or not a horseshoe dynamics still exists for those smaller planets
 for which the U-turn drift rate is smaller than the typical amplitude
 of the gas turbulent velocity fluctuations.  This regime deserves a
 dedicated investigation that goes beyond the scope of this paper.

\subsection{Averaged torque density distribution}
\label{sec:tqdistri}
The results of simulations presented in Sect.~\ref{sec:cavity}
and~\ref{sec:power} support the existence of an unsaturated corotation
torque in weakly magnetized turbulent discs. The running time averaged
torques obtained with the NIRVANA and RAMSES codes are in very good
agreement overall. Given the small scales we are interested in, this
is reassuring as to the robustness of our results.

 For the two disc models in Sect.~\ref{sec:power}, we find a decent
 agreement overall between the steady running time-averages of the
 tidal torque in the turbulent and viscous simulations. In Model 1,
 the torque relative difference between turbulent and viscous models
 is less than $30\%$ for both codes. In Model 2, however, it is less
 than $60\%$. To provide more insight into these torque differences,
 we compare in Fig.~\ref{fig:tqprof} the time-averaged torque density
 profiles obtained for both models (depicted by plus symbols), with
 the torque density profiles of the viscous and inviscid laminar disc
 models (star and diamond symbols, respectively). Results with FARGO
 and NIRVANA are shown in the left panel, those with RAMSES in the
 right panel. The torque profiles of the MHD runs are time-averaged
 over 100 orbits. In each plot, the approximate location of the
 separatrices of the planet's horseshoe region is indicated by dashed
 lines.

 We first compare the torque density distributions obtained in the
 laminar disc models. In Model 1, the torque distributions of the
 viscous and inviscid runs mostly differ within the planet's horseshoe
 region.  This difference highlights the positive unsaturated
 corotation torque of the viscous disc model. Slight differences can
 be noticed between the corotation torque distributions of the FARGO
 and RAMSES simulations. In Model 2, where there is no corotation
 torque, the torque distributions of the viscous and inviscid
 simulations are in very good agreement. Their tiny differences show
 the influence of viscous diffusion on the radial profile of the
 differential Lindblad torque.
 
 We now compare the averaged torque distribution of the MHD runs (plus
 symbols) with the torque distribution of the viscous runs (star
 symbols). Outside the planet's (mean) horseshoe region, both
 distributions are found to be in overall good agreement. This is a
 clear demonstration that the differential Lindblad torque remains
 essentially unchanged by the full development of MHD
 turbulence. Inside the planet's horseshoe region, however,
 differences are significant. In Model 2, the turbulent torque
 distribution takes more negative values over most of the horseshoe
 region, except near the outer separatrix, where it takes more
 positive values. Note the excellent agreement between the turbulent
 torque distributions of the NIRVANA and RAMSES runs. In contrast, the
 turbulent torque distribution obtained in Model 1 tends to take more
 positive values over the horseshoe region, except near the inner
 separatrix, although this trend is less evident in the RAMSES run.
 These torque differences suggest the existence of an additional
 torque in weakly magnetized turbulent discs that originates from the
 planet's horseshoe region. The distribution of this additional
 torque, in particular its asymmetry, accounts for the torque
 differences between the MHD turbulent and viscous simulations
 underlined in Sect.~\ref{sec:model1} and~\ref{sec:model2}.
 
 It is quite surprising that the density distribution of the
 additional torque is not symmetric with respect to the location of
 the planet, or of the planet's corotation radius. This offset is
 obtained with our two codes, but it differs in Models 1 and
 2. Notwithstanding the offset, both models point out that the
 additional torque tends to take negative values in the inner half of
 the horseshoe region, and positive values in the outer half.  This
 suggests that the additional torque is probably not powered by
 magnetic resonances. As shown by \cite{Terquem03}, the torques at the
 magnetic and Lindblad resonances have the same sign. The torque
 exerted by the disc on the planet near the inner (outer) magnetic
 resonance is therefore positive (negative) \citep[see also][their Fig
 6]{Fromangetal05}. The torque distribution of this so-called magnetic
 torque, and of the additional torque in our magnetized turbulent disc
 models, therefore have opposite signs. We also point out that
 magnetic resonances are barely resolved in our setup. Their location,
 $R_{\rm m}$, as given by linear theory \citep{Terquem03}, satisfies
 $|R_{\rm m} - R_{\rm p}| = (2H_{\rm p}/3) (1+\beta)^{-1/2}$, with
 $\beta$ the effective plasma parameter evaluated from the average of
 the magnetic field squared. For the NIRVANA runs, we find $\beta
 \approx 40$, and $|R_{\rm m} - R_{\rm p}| \approx 0.03$ is scarcely
 larger than our mesh size ($\approx 0.02$).
   
 It is possible that the additional torque found in our magnetized
 turbulent disc models is an additional corotation torque. Neglecting
 turbulence for the time being, in a magnetized disc with a purely
 toroidal magnetic field the magnetic tension tries to prevent radial
 motion associated with horseshoe U-turns imposed by the planet's
 gravity, and attempts to force fluid element trajectories to follow
 field lines. When the magnetic field is strong enough, it is
 conceivable that magnetic tension can prevent horseshoe U-turns,
 cancelling the corotation torque. This might explain why the torque
 distribution in the 2D MHD simulations of \cite{Fromangetal05}, where
 $\beta=2$, reveals no corotation torque, but instead a magnetic
 torque with properties similar to those predicted by linear
 theory. However, a weak magnetic field, such as the one we consider
 here, cannot prevent horseshoe U-turns, as we have demonstrated in
 Sect.~\ref{sec:hs}. In this case, we speculate that magnetic tension
 will lead to asymmetric horseshoe U-turns, with downstream
 streamlines being closer to the corotation radius than the upstream
 streamlines. This is equivalent to saying that fluid elements in
 weakly magnetized discs will experience smaller changes to their
 angular momenta during horseshoe U-turns than they will in
 non-magnetized discs, with a consequence being that the density near
 the downstream separatrices will be reduced.  Magnetic tension will
 thus reduce the positive corotation torque on the planet arising from
 fluid elements undergoing inward horseshoe U-turns, and will reduce
 the magnitude of the negative corotation torque arising from fluid
 elements undergoing outward horseshoe U-turns. This predicted
 behaviour is in broad agreement with the results presented in
 Fig.~\ref{fig:tqprof}, provided we ignore the fact that the
 modification to the corotation torque observed in the simulations is
 not quite symmetric about the corotation radius. The origin of this
 offset about the corotation radius is not understood at present. A
 more detailed investigation of horseshoe dynamics in the presence of
 a weak toroidal magnetic field and MHD turbulence go beyond the scope
 of this work, however, and will be presented in a forthcoming
 publication.
 
 We finally comment that the results of this study have been obtained
 with rather large values of the disc aspect ratio ($h_{\rm p} = 0.1$)
 and of the planet mass ($M_{\rm p} = 3\times 10^{-4} M_{\star}$).
 Although this planet mass is relevant to type I migration (see
 Sect.~\ref{sec:planetparam}), it is \emph{a priori} not
 straightforward to adapt our results to smaller, more canonical
 values of $h_{\rm p}$ and $M_{\rm p}$, or to different values of the
 alpha viscosity parameter associated with the turbulent stress. This
 requires knowledge of the properties of the \emph{total} corotation
 torque in laminar magnetized discs, and whether or not they are
 altered by MHD turbulence.

\subsection{Mass-adding procedure}
\label{sec:addmass} The MHD simulations presented in this study were
carried out using the mass-adding procedure described in
Sect.~\ref{sec:physmodel}.  This procedure, which restores the initial
density profile on a timescale set to 20 orbital periods, aims at
obtaining a steady-state density profile, mass being otherwise
continuously relocated by the turbulence towards the radial boundaries
of the computational domain.  Not including this procedure would make
both the density profile and the running time-averaged tidal torque
non-stationary quantities.  In the simulations presented in
Sect.~\ref{sec:power}, the choice for a restoring timescale of 20
orbital periods helps obtain a time-averaged density profile close to
the initial profile, and which can be approximated as a power-law
function of radius over a rather large radial extent about the
planet's orbital radius (see Fig.~\ref{fig:rtadens_q1}). This has
enabled us to compare the torques of our simulations with analytical
torque formulae derived in disc models with power-law density
profiles. A different restoring timescale would result in a different
density profile in a steady-state, and therefore in a different tidal
torque. Although not presented here, we performed another MHD
simulation with the parameters of Model 1 and a restoring timescale
increased to 40 orbital periods. This additional run was performed
with the NIRVANA code. We also carried out an additional 2D viscous
simulation using the FARGO code, where the initial profiles of the
density and of the viscous alpha parameter correspond to their
time-averaged counterpart in the MHD simulation (this is the same
approach as that adopted in Sect.~\ref{sec:power}). The agreement
between the laminar and turbulent simulations is very similar to that
obtained with a restoring timescale of 20 orbital periods: the
time-averaged turbulent and laminar torques agree to within $\sim
10\%$, and the torque density distributions show the same features as
in the top-right panel of Fig.~\ref{fig:tqprof}. This agreement makes
us confident that the mass-adding procedure does not introduce any
other artifacts on the time-averaged tidal torque than the one
naturally arising from changing the density profile in a steady-state
by choosing a different restoring timescale. We finally stress that
this procedure should not be used in turbulent disc models involving a
gap-opening planet.
 
\section{Conclusions}
\label{sec:conclu}
In this paper we have investigated the tidal torque between a planet
on a fixed circular orbit and its nascent protoplanetary disc,
assuming that the latter sustains turbulence driven by the non-linear
development of the MRI. Our main aim has been to investigate the
properties of the corotation torque in such turbulent discs.  In this
initial study, we adopted simple 3D magnetized disc models with a
locally isothermal equation of state, and neglected vertical
stratification and non-ideal MHD effects.  We performed simulations
using two different MHD codes (NIRVANA and RAMSES) over time spans of
several hundred planet orbits.

In one set of simulations, we considered the torque experienced by a
planet orbiting at the outer edge of an inner disc cavity, in a region
with a strong, positive density gradient.  The running time-averaged
torque in this case was found to be strongly positive over several
hundred planet orbits, in good agreement with the results of laminar
viscous disc simulations whose viscous stresses were similar to that
generated by the MHD turbulence. This positive torque strongly
suggests the existence of an unsaturated corotation torque in weakly
magnetized turbulent disc models. It implies that the presence of an
inner cavity can stall type I migration, such that a planet trap can
operate in turbulent discs.

We also performed simulations of two disc models with power-law
initial density profiles. We found that the running time-averaged
torque of the turbulent disc models reached a stationary value after
100 to 200 planet orbits, with decent agreement between it and a
laminar viscous disc model with similar viscous alpha parameter at the
planet location. The relative difference between the averaged torques
in turbulent and laminar disc simulations ranges from $20\%$ to
$60\%$, depending on the code and the disc model. We find remarkably
good overall agreement between the results of the two codes, and given
the scales of interest (the half-width of the planet's horseshoe
region is a fraction of the pressure scale height), this agreement
highlights the robustness of our results.

Analysis of the velocity fields in our simulations shows the existence
of horseshoe dynamics in turbulent discs with a weak toroidal magnetic
field. The shape and width of the time-averaged horseshoe region much
resemble those of laminar viscous discs with similar viscosity at the
planet's location. Furthermore, close inspection of the averaged
torque density distributions demonstrates that the differential
Lindblad torque takes very similar values in turbulent and laminar
viscous disc models, and there exists an unsaturated corotation torque
in weakly magnetized turbulent discs. This analysis, however, also
indicates the existence of an additional torque with respect to the
non-magnetic case, which originates from the planet's horseshoe
region. We argue that this additional torque is likely to be an
additional corotation torque, possibly arising from the effects of
magnetic tension.

Although we have demonstrated conclusively the existence of
unsaturated corotation torques in weakly magnetized turbulent discs,
our study contains a number of simplifications that need to be
addressed. We have considered disc and planet parameters such that the
horseshoe U-turn drift rate exceeds the amplitude of the turbulent
velocity fluctuations. In the opposite case, the existence of
horseshoe dynamics and a corotation torque is unknown. We have
neglected vertical stratification and non-ideal MHD effects such as
Ohmic resistivity, and it is fully expected that inclusion of these
effects will modify the results presented here. We have also kept the
planet on a fixed circular orbit and simply measured the tidal torque
as a time series. In reality the planet will migrate and experience
evolution in its eccentricity and inclination as a result of
interaction with the turbulent disc. We will address these issues in a
series of forthcoming publications, along with an in-depth
investigation of the additional corotation torque in weakly magnetized
discs that our simulations have uncovered.

\section*{ACKNOWLEDGMENTS}
CB acknowledges support from a Herchel Smith Postdoctoral Fellowship.
SF acknowledges the help of Edouard Audit in extending RAMSES to
cylindrical coordinates systems. The simulations performed with RAMSES
were granted access to the HPC resources of CCRT and CINES under the
allocation x2010042231 made by GENCI (Grand Equipement National de
Calcul Intensif). Simulations with NIRVANA and FARGO were carried out
on the Pleiades Cluster at U.C. Santa Cruz, on the Darwin
Supercomputer of the University of Cambridge High Performance
Computing Service (http://www.hpc.cam.ac.uk), provided by Dell
Inc. using Strategic Research Infrastructure Funding from the Higher
Education Funding Council for England. NIRVANA simulations were also
performed on the QMUL HPC facilities purchased under the SRIF/CIF
initiatives. This research was initiated during the Isaac Newton
Institute programme `Dynamics of Discs and Planets'.  We thank
J. Guilet and J. Papaloizou for useful discussions, and A. Uribe, the
referee, for a prompt and useful report.


\begin{thebibliography}{42}
\expandafter\ifx\csname natexlab\endcsname\relax\def\natexlab#1{#1}\fi

\bibitem[{{Baruteau} \& {Lin}(2010)}]{bl10}
{Baruteau}, C. \& {Lin}, D.~N.~C. 2010, \apj, 709, 759

\bibitem[{{Baruteau} \& {Masset}(2008)}]{bm08a}
{Baruteau}, C. \& {Masset}, F. 2008, \apj, 672, 1054

\bibitem[{{Crida} \& {Morbidelli}(2007)}]{cm07}
{Crida}, A. \& {Morbidelli}, A. 2007, \mnras, 377, 1324

\bibitem[{{Crida} {et~al.}(2006){Crida}, {Morbidelli}, \& {Masset}}]{crida06}
{Crida}, A., {Morbidelli}, A., \& {Masset}, F. 2006, Icarus, 181, 587

\bibitem[{{de Val-Borro} {et~al.}(2006){de Val-Borro}, {Edgar}, {Artymowicz},
  {Ciecielag}, {Cresswell}, {D'Angelo}, {Delgado-Donate}, {Dirksen}, {Fromang},
  {Gawryszczak}, {Klahr}, {Kley}, {Lyra}, {Masset}, {Mellema}, {Nelson},
  {Paardekooper}, {Peplinski}, {Pierens}, {Plewa}, {Rice}, {Sch{\"a}fer}, \&
  {Speith}}]{valborro06}
{de Val-Borro}, M., {Edgar}, R.~G., {Artymowicz}, P., {et~al.} 2006, \mnras,
  695

\bibitem[{{Fromang} {et~al.}(2006){Fromang}, {Hennebelle}, \&
  {Teyssier}}]{fromangetal06}
{Fromang}, S., {Hennebelle}, P., \& {Teyssier}, R. 2006, \aap, 457, 371

\bibitem[{{Fromang} {et~al.}(2005){Fromang}, {Terquem}, \&
  {Nelson}}]{Fromangetal05}
{Fromang}, S., {Terquem}, C., \& {Nelson}, R.~P. 2005, \mnras, 363, 943

\bibitem[{{Hasegawa} \& {Pudritz}(2010)}]{hp10}
{Hasegawa}, Y. \& {Pudritz}, R.~E. 2010, \apjl, 710, L167

\bibitem[{{Ida} \& {Lin}(2008{\natexlab{a}})}]{IdaLin4}
{Ida}, S. \& {Lin}, D.~N.~C. 2008{\natexlab{a}}, \apj, 673, 487

\bibitem[{{Ida} \& {Lin}(2008{\natexlab{b}})}]{IdaLin5}
{Ida}, S. \& {Lin}, D.~N.~C. 2008{\natexlab{b}}, \apj, 685, 584

\bibitem[{{Laughlin} {et~al.}(2004){Laughlin}, {Steinacker}, \&
  {Adams}}]{lsa04}
{Laughlin}, G., {Steinacker}, A., \& {Adams}, F.~C. 2004, \apj, 608, 489

\bibitem[{{Lin} \& {Papaloizou}(1986)}]{lp86}
{Lin}, D.~N.~C. \& {Papaloizou}, J. 1986, \apj, 309, 846

\bibitem[{{Masset}(2000{\natexlab{a}})}]{fargo1}
{Masset}, F. 2000{\natexlab{a}}, \aaps, 141, 165

\bibitem[{{Masset}(2000{\natexlab{b}})}]{fargo2}
{Masset}, F.~S. 2000{\natexlab{b}}, in Astronomical Society of the Pacific
  Conference Series, Vol. 219, Disks, Planetesimals, and Planets, ed.
  G.~{Garz{\'o}n}, C.~{Eiroa}, D.~{de Winter}, \& T.~J. {Mahoney}, 75--+

\bibitem[{{Masset}(2001)}]{masset01}
{Masset}, F.~S. 2001, \apj, 558, 453

\bibitem[{{Masset} \& {Casoli}(2010)}]{mc10}
{Masset}, F.~S. \& {Casoli}, J. 2010, \apj, 723, 1393

\bibitem[{{Masset} {et~al.}(2006{\natexlab{a}}){Masset}, {D'Angelo}, \&
  {Kley}}]{mak2006}
{Masset}, F.~S., {D'Angelo}, G., \& {Kley}, W. 2006{\natexlab{a}}, \apj, 652,
  730

\bibitem[{{Masset} {et~al.}(2006{\natexlab{b}}){Masset}, {Morbidelli}, {Crida},
  \& {Ferreira}}]{masset06a}
{Masset}, F.~S., {Morbidelli}, A., {Crida}, A., \& {Ferreira}, J.
  2006{\natexlab{b}}, \apj, 642, 478

\bibitem[{{Masset} \& {Papaloizou}(2003)}]{mp03}
{Masset}, F.~S. \& {Papaloizou}, J.~C.~B. 2003, \apj, 588, 494

\bibitem[{{Menou} \& {Goodman}(2004)}]{mg2004}
{Menou}, K. \& {Goodman}, J. 2004, \apj, 606, 520

\bibitem[{{Morbidelli} {et~al.}(2008){Morbidelli}, {Crida}, {Masset}, \&
  {Nelson}}]{morby08}
{Morbidelli}, A., {Crida}, A., {Masset}, F., \& {Nelson}, R.~P. 2008, \aap,
  478, 929

\bibitem[{{Mordasini} {et~al.}(2009){Mordasini}, {Alibert}, {Benz}, \&
  {Naef}}]{Mordasini09b}
{Mordasini}, C., {Alibert}, Y., {Benz}, W., \& {Naef}, D. 2009, \aap, 501, 1161

\bibitem[{{Muto} \& {Inutsuka}(2009)}]{Muto09}
{Muto}, T. \& {Inutsuka}, S. 2009, \apj, 701, 18

\bibitem[{{Muto} {et~al.}(2008){Muto}, {Machida}, \& {Inutsuka}}]{Muto08}
{Muto}, T., {Machida}, M.~N., \& {Inutsuka}, S.-i. 2008, \apj, 679, 813

\bibitem[{{Nelson}(2005)}]{nelson05}
{Nelson}, R.~P. 2005, \aap, 443, 1067

\bibitem[{{Nelson} \& {Gressel}(2010)}]{ng10}
{Nelson}, R.~P. \& {Gressel}, O. 2010, \mnras, 409, 639

\bibitem[{{Nelson} \& {Papaloizou}(2003)}]{qmwmhd2}
{Nelson}, R.~P. \& {Papaloizou}, J.~C.~B. 2003, \mnras, 339, 993

\bibitem[{{Nelson} \& {Papaloizou}(2004)}]{np2004}
{Nelson}, R.~P. \& {Papaloizou}, J.~C.~B. 2004, \mnras, 350, 849

\bibitem[{{Paardekooper} {et~al.}(2010){Paardekooper}, {Baruteau}, {Crida}, \&
  {Kley}}]{pbck10}
{Paardekooper}, S., {Baruteau}, C., {Crida}, A., \& {Kley}, W. 2010, \mnras,
  401, 1950

\bibitem[{{Paardekooper} {et~al.}(2011){Paardekooper}, {Baruteau}, \&
  {Kley}}]{pbk11}
{Paardekooper}, S., {Baruteau}, C., \& {Kley}, W. 2011, \mnras, 410, 293

\bibitem[{{Paardekooper} \& {Papaloizou}(2008)}]{pp08}
{Paardekooper}, S.-J. \& {Papaloizou}, J.~C.~B. 2008, \aap, 485, 877

\bibitem[{{Paardekooper} \& {Papaloizou}(2009)}]{pp09a}
{Paardekooper}, S.-J. \& {Papaloizou}, J.~C.~B. 2009, \mnras, 394, 2283

\bibitem[{{Schlaufman} {et~al.}(2009){Schlaufman}, {Lin}, \& {Ida}}]{sli09}
{Schlaufman}, K.~C., {Lin}, D.~N.~C., \& {Ida}, S. 2009, \apj, 691, 1322

\bibitem[{{Simon} {et~al.}(2009){Simon}, {Hawley}, \& {Beckwith}}]{Simon09}
{Simon}, J.~B., {Hawley}, J.~F., \& {Beckwith}, K. 2009, \apj, 690, 974

\bibitem[{{Stone} \& {Norman}(1992)}]{zeus}
{Stone}, J.~M. \& {Norman}, M.~L. 1992, \apjs, 80, 753

\bibitem[{{Tanaka} {et~al.}(2002){Tanaka}, {Takeuchi}, \& {Ward}}]{tanaka2002}
{Tanaka}, H., {Takeuchi}, T., \& {Ward}, W.~R. 2002, \apj, 565, 1257

\bibitem[{{Terquem}(2003)}]{Terquem03}
{Terquem}, C.~E.~J.~M.~L.~J. 2003, \mnras, 341, 1157

\bibitem[{{Teyssier}(2002)}]{teyssier02}
{Teyssier}, R. 2002, \aap, 385, 337

\bibitem[{{Uribe} {et~al.}(2011){Uribe}, {Klahr}, {Flock}, \&
  {Henning}}]{Uribe11}
{Uribe}, A., {Klahr}, H., {Flock}, M., \& {Henning}, T. 2011, ArXiv e-prints

\bibitem[{{Ward}(1991)}]{wlpi91}
{Ward}, W.~R. 1991, in Lunar and Planetary Institute Conference Abstracts,
  1463--+

\bibitem[{{Ward}(1997)}]{w97}
{Ward}, W.~R. 1997, Icarus, 126, 261

\bibitem[{{Ziegler} \& {Yorke}(1997)}]{nirvana1}
{Ziegler}, U. \& {Yorke}, H.~W. 1997, Computer Physics Communications, 101, 54

\end{thebibliography}
\end{document}